\documentclass[acmsmall,screen,nonacm]{acmart}

\usepackage{amsmath}
\usepackage{algorithmic}
\usepackage{graphicx}
\usepackage{textcomp}
\usepackage{xcolor}

\usepackage{listings}
\usepackage{todonotes}
\usepackage{minted}
\usepackage{booktabs}
\usepackage{subcaption}
\usepackage{adjustbox}
\usepackage{tabularx}
\usepackage{array}

\usepackage[inline]{enumitem}
\setitemize{noitemsep,topsep=1pt,parsep=1pt,partopsep=1pt}

\usepackage[most]{tcolorbox}
\tcbset{
  colback=gray!10,    
  colframe=gray!80,   
  boxrule=0.8pt,      
  arc=3pt,            
  left=6pt, right=6pt, top=6pt, bottom=6pt,
  fonttitle=\bfseries,
}

\newcommand{\rep}[2]{\mbox{\texttt{#1} (#2\%)}}
\newcolumntype{L}{>{\raggedright\arraybackslash}X}

\definecolor{codegreen}{rgb}{0,0.6,0}
\lstdefinestyle{CoqStyle}{
    backgroundcolor=\color{white},   
    commentstyle=\color{codegreen},
    keywordstyle=\color{blue},
    numberstyle=\tiny\color{black},
    stringstyle=\color{red},
    basicstyle=\ttfamily\scriptsize,
    morekeywords={Require,Import,Proof,Qed,Lemma,Theorem,Admitted},
    breakatwhitespace=false,         
    breaklines=true,                 
    captionpos=b,                    
    keepspaces=true,                 
    numbers=left,                    
    xleftmargin=0ex, 
    numbersep=5pt,                  
    showspaces=false,                
    showstringspaces=false,
    showtabs=false,                  
    tabsize=2
}

\def\BibTeX{{\rm B\kern-.05em{\sc i\kern-.025em b}\kern-.08em
    T\kern-.1667em\lower.7ex\hbox{E}\kern-.125emX}}

\lstdefinelanguage{Coq}{
  morekeywords={
    Theorem,Lemma,Proof,Qed,Definition,Fixpoint,
    Inductive,match,with,end,fun,forall,exists
  },
  sensitive=true,
  morecomment=[n]{(*}{*)},
  morestring=[b]"
}

\newcommand{\JFF}[1]{\textcolor{blue}{#1}\GenericWarning{}{LaTeX Warning: JFF: #1}}\newcommand\jff\JFF

\newcommand{\RQAnswer}[2]{%
\begin{tcolorbox}[
    enhanced,
    breakable,
    colback=gray!5,
    colframe=black!40,
    boxrule=0.3pt,
    borderline west={2pt}{0pt}{black!80},
    left=2mm,
    right=2mm,
    top=1mm,
    bottom=1mm
]
\textbf{#1} #2
\end{tcolorbox}
}

\AtBeginDocument{%
  \providecommand\BibTeX{{%
    Bib\TeX}}}

\setcopyright{none}
\begin{document}
\pagestyle{plain}

\author{Martim Afonso}
\orcid{XXXX-XXXX-XXXX-XXXX}
\affiliation{%
  \institution{INESC-ID \& IST, University of Lisbon}
  \city{Lisbon}
  \country{Portugal}
}
\affiliation{%
  \institution{Politecnico di Torino}
  \city{Turin}
  \country{Italy}
}
\email{martim.afonso@tecnico.ulisboa.pt}

\author{Inês Lynce}
\orcid{XXXX-XXXX-XXXX-XXXX}
\affiliation{%
  \institution{INESC-ID \& IST, University of Lisbon}
  \city{Lisbon}
  \country{Portugal}
}
\email{ines.lynce@tecnico.ulisboa.pt}

\author{Maurizio Morisio}
\orcid{XXXX-XXXX-XXXX-XXXX}
\affiliation{%
  \institution{Politecnico di Torino}
  \city{Turin}
  \country{Italy}
}
\email{maurizio.morisio@polito.it}

\author{João F. Ferreira}
\orcid{0000-0002-6612-9013}
\affiliation{%
  \institution{INESC-ID \& Faculty of Engineering, University of Porto}
  \city{Porto}
  \country{Portugal}
}
\email{joao@joaoff.com}

\title[\toolname]{\toolname: Continuous Reasoning for Automated Software Verification}

\renewcommand{\shortauthors}{}
\newcommand{\toolname}{\mbox{ProofCoder}}
\begin{abstract}
Formal verification using proof assistants provides strong correctness guarantees for software, but constructing proofs remains labor-intensive and requires significant expertise. Recent approaches based on large language models (LLMs) have shown promise in automating proof synthesis. However, these approaches typically constrain the model to a rigid tactic-generation loop, in which the surrounding system controls premise retrieval, proof-state management, and backtracking.

We introduce \toolname, an agentic proof-synthesis system for Rocq that approaches theorem proving from a coding-agent perspective. Rather than generating tactics within a fixed loop, \toolname\ reasons continuously throughout proof search and uses proof-assistant feedback to decide when to gather additional information, inspect proof states, revise previously generated proof fragments, or backtrack to an earlier point in the proof. This allows it to iteratively construct, repair, and refine proofs, at either tactic-level or whole-proof granularity.

We evaluate \toolname\ on 280 challenging verification tasks from the AutoRocq Benchmark. On the best-performing configurations, \toolname\ increases proof success from 47\% to 60\% over the state-of-the-art AutoRocq system, corresponding to a 27.7\% relative improvement at approximately half the cost.
Our ablations identify continuous reasoning as a key design choice: without it, the agent solves fewer lemmas at substantially higher cost, backtracks four to six times more often, and more frequently discards its own revisions. Whole-proof and tactic-level generation achieve similar success rates but follow distinct search and repair strategies.
These results suggest that effective automated software verification requires more than improved tactic selection: continuous reasoning and feedback-guided proof revision are critical ingredients for robust proof synthesis.
\end{abstract}




\maketitle
\thispagestyle{plain}

\section{Introduction}

As large language models (LLMs) are increasingly used for code generation, there is a growing need to verify generated software for bugs, vulnerabilities, and violations of intended specifications~\cite{software_quality}. Formal verification offers a principled way to establish strong correctness guarantees and has been used in several real-world systems~\cite{sel4, compcert, rdbms, fv_survey}. In many verification workflows, programs and specifications are translated into logical proof obligations that can be discharged using interactive theorem provers (ITPs), such as Lean~\cite{lean} and Rocq~\cite{coq}.
However, formal verification through ITPs remains expensive: users must formulate appropriate specifications and construct proof scripts accepted by the proof assistant, a process that is time-consuming and requires substantial expertise. Moreover, changes to the program or its specifications can invalidate existing proofs, requiring proofs to be adapted or reconstructed as the software evolves.

LLM-based proof synthesis is a promising direction for reducing the burden of interactive theorem proving. Compared with code generation, proof generation has a distinctive advantage: candidate proof scripts are checked by the proof assistant. As a result, hallucinated or nonsensical model outputs do not silently compromise correctness, as they are rejected by the proof assistant and can be treated as failed attempts. This makes theorem proving well suited to LLM-based approaches, combining strong, executable feedback with the soundness guarantees of the underlying prover.

A common design in prior LLM-based proof systems is to place the model within a controller-driven search loop, where the surrounding algorithm determines when and how to retrieve context, backtrack, or repair~\cite{cobblestone, copra, rango, palm, prooverbot9001, qedcartographer, tactok}. 
AutoRocq~\cite{autorocq} moves toward an agentic formulation of software verification, showing that giving the model greater control over proof search can substantially improve over earlier controller-driven approaches.
Instantiated in Rocq, it targets program-verification obligations and allows the LLM to select among tools for planning, tactic generation, and querying the verification environment. Nevertheless, its autonomy is bounded by a semi-constrained protocol that restricts tool use and guides subsequent actions.

In this work, we introduce \toolname, an agentic proof-synthesis system inspired by modern coding agents~\cite{sweagent}. Rather than following a fixed or highly structured interaction loop, \toolname\ gives the agent greater control over the proof process: it can inspect previous proof states, query the environment, read the source file, edit the proof script, and continuously reason about how to combine these actions, using ITP feedback to guide proof construction and repair.
This design lets us study whether a coding-agent-style interaction model is better suited to interactive theorem proving than more constrained tactic-generation loops, and how restrictions on tools, generation granularity, and interaction structure affect proving performance.

We evaluate \toolname\ on 280 challenging verification lemmas from the AutoRocq Benchmark, drawn from SV-COMP~\cite{svcomp} and the Linux kernel (Verker). Our primary comparison is with AutoRocq~\cite{autorocq}, the state of the art in agentic Rocq-based software verification. We additionally compare against Rango~\cite{rango}, PALM~\cite{palm}, COPRA~\cite{copra}, QEDCartographer~\cite{qedcartographer}, and Proverbot9001~\cite{prooverbot9001}.

Across both benchmarks and two different models, \toolname\ outperforms the state of the art, increasing proof success from 47\% to 60\% in the best-performing configuration while also reducing inference cost by approximately half. We further conduct a series of ablations showing that continuous reasoning and localized repair are key to these gains.


In summary, the main \textbf{contributions} of this paper are:
\begin{itemize}
    \item A new approach to agentic theorem proving, inspired by coding agents, that uses continuous reasoning and a more flexible agentic loop to outperform the state of the art.
    \item Several ablations and analyses that quantify the impact of different constraints on agent performance and behavior.
    \item A pipeline for theorem proving in Rocq with a simple and adaptable Model Context Protocol. 
\end{itemize}

\section{Motivating Example}

We illustrate the behavior of \toolname\ with the loop-invariant preservation property shown in Listing~\ref{lst-motivating-initial}, adapted from the SV-COMP benchmark. This example is representative of the verification lemmas and proof engineering tasks encountered in software verification with interactive proof assistants such as Rocq, particularly those generated by tools such as Frama-C~\cite{kirchner2015frama}.
\begin{lstlisting}[
style=CoqStyle,
numbers=none,
caption={Abridged loop-invariant preservation lemma.},
label={lst-motivating-initial}
]
Theorem wp_goal :
forall (t : addr -> int) (i i1 : int) (a : addr),
 let x := t (shift a i1) in
 0 < i1 -> 0 < i -> i1 <= i ->
 (* region and bound hypotheses omitted *)
 is_sint32 (t (shift a i)) ->
 (forall i2,
   let x1 := t (shift a i2) in 0 < i2 -> i2 < i ->
  -2147483648 <= x1 <= 2147483647) -> 
  -2147483648 <= x <= 2147483647.
\end{lstlisting}
\vskip 0.3em
\noindent\noindent
\emph{\textbf{Intuition.}}
The loop invariant states that every array element processed so far (i.e., every index \texttt{i2 < i}) is a valid signed 32-bit integer (i.e., within the bounds $-2147483648$ and $2147483647$). The goal is to prove that the same property holds for an arbitrary index \texttt{i1} satisfying \texttt{i1 <= i}. 
\vskip 0.3em
\noindent\noindent
\emph{\textbf{Challenge.}} 
The proof obligation is simple at a high level, but it requires careful interaction with the evolving proof state. The invariant hypothesis applies only to indices strictly smaller than \texttt{i}, while the context gives only \texttt{i1 <= i}. Thus, applying the invariant directly is insufficient: the proof must distinguish between the cases \texttt{i1 < i} and \texttt{i1 = i}. In the first case, the invariant applies directly. In the second case, the goal reduces to the existing hypothesis \texttt{is\_sint32(t\ (shift\ a\ i))}, after unfolding the definition of \texttt{is\_sint32}.
\vskip 0.3em
\noindent\noindent
\emph{\textbf{AutoRocq behavior.}}
AutoRocq, the state of the art in agentic Rocq-based software verification, identifies the main proof structure early: it observes that the proof must distinguish between the cases \texttt{i1 < i} and \texttt{i1 = i}. This is the right high-level decomposition, since the loop invariant can only be applied to indices strictly smaller than \texttt{i}, while the context provides only \texttt{i1 <= i}. However, after reaching the boundary case \texttt{i1 = i}, AutoRocq fails to turn this decomposition into a completed proof.

In this branch, rewriting \texttt{i1} to \texttt{i} reduces the goal to an arithmetic bound on \texttt{t\ (shift\ a\ i)}. The context already contains the hypothesis

\begin{lstlisting}[
style=CoqStyle,
xleftmargin=5.0ex,
frame=none,
numbers=none]
H_sint32_i_val : is_sint32 (t (shift a i))
\end{lstlisting}
which is sufficient after unfolding the definition of \texttt{is\_sint32}. 
AutoRocq retrieves this definition via \texttt{Print}, but
does not unfold it, instead exploring transformations of the address expression
such as rewriting and unfolding \texttt{shift}, and attempting to prove equalities between terms. These steps move the proof away from the fact needed to close the branch.

Thus, AutoRocq's failure is not due to missing the global proof idea. It finds the same case split that is needed for the proof. The failure occurs at the level of local proof-state interpretation: once the proof reaches the branch where \texttt{i1~=~i}, AutoRocq does not recognize that the remaining goal is an unfolded form of an existing \texttt{is\_sint32} hypothesis. The result is a long sequence of unproductive repairs, eventually exhausting the token budget without completing the proof.
\vskip 0.3em
\noindent\noindent
\emph{\textbf{How \toolname\ succeeds.}}
\toolname\ begins from the natural but incomplete strategy of applying the loop invariant directly at index \texttt{i1}. Initially, it tries the following:

\begin{lstlisting}[
style=CoqStyle,
numbers=none]
intros ... H_forall.
unfold x.
apply H_forall.
- assumption.
- lia.
\end{lstlisting}
This attempt fails because the remaining subgoal is \texttt{i1 < i}. The proof state therefore exposes the missing condition: the context contains \texttt{i1 <= i}, but the invariant requires a strict inequality. \toolname\ receives this feedback and, through continuous reasoning, revises its strategy rather than abandoning the proof or continuing with unrelated rewrites:

\newtcolorbox{reasoningbox}{
  colback=blue!3,
  colframe=blue!45!black,
  boxrule=0.5pt,
  arc=2pt,
  left=6pt,
  right=6pt,
  top=5pt,
  bottom=5pt,
  fonttitle=\bfseries,
  title=Agent reasoning,
  before skip=6pt,
  after skip=6pt
}

\newenvironment{agentreasoning}
{\par\vspace{0.25em}\noindent\begingroup\footnotesize\itshape\leftskip=1em}
{\par\endgroup\vspace{0.25em}}

\begin{agentreasoning}
The issue is that we need to show \texttt{i1 < i}, but we only have 
\texttt{i1 <= i}. Let me reconsider the proof strategy. The universal quantifier requires \texttt{i2 < i}, but we only know \texttt{i1 <= i}. I need to handle the case where \texttt{i1 = i} separately.    
\end{agentreasoning}
\noindent It backtracks to the beginning of the proof fragment and introduces the needed case split:

\begin{lstlisting}[
style=CoqStyle,
numbers=none,
escapeinside={(*@}{@*)}]
intros ... H_forall.
unfold x.
assert (i1 < i \/ i1 = i) as [H_lt | H_eq] by lia.
\end{lstlisting}
This repair preserves the useful prefix of the proof, namely the introduced variables and the unfolding of \texttt{x}, while replacing only the incorrect attempt to apply the invariant unconditionally. In the \texttt{i1 < i} branch, \toolname\ reuses the original idea and applies \texttt{H\_forall} successfully:

\begin{lstlisting}[
style=CoqStyle,
numbers=none]
apply H_forall.
assumption.
\end{lstlisting}
The remaining branch is the boundary case \texttt{i1=i}. Rewriting with \texttt{H\_eq} gives:

\begin{lstlisting}[
style=CoqStyle,
xleftmargin=5.0ex,
frame=none,
numbers=none]
-2147483648 <= t (shift a i) <= 2147483647
\end{lstlisting}
Here, \texttt{assumption} fails because the matching hypothesis uses the defined predicate:

\begin{lstlisting}[
style=CoqStyle,
xleftmargin=5.0ex,
frame=none,
numbers=none]
H_sint32_i_val : is_sint32 (t (shift a i))
\end{lstlisting}
\toolname\ queries the definition of \texttt{is\_sint32}, learns that it expands to \texttt{-2147483648 <= x < 2147483648}, 
and, through its continuous reasoning process, comes up with:

\begin{agentreasoning}
The goal requires \texttt{... <= 2147483647} while \texttt{is\_sint32} uses \texttt{< 2147483648}. These are equivalent for integers. Let me apply \texttt{H\_sint32\_i\_val} and use \texttt{lia}.
\end{agentreasoning}
\noindent It then locally replaces the failed \texttt{assumption} step with:

\begin{lstlisting}[
style=CoqStyle,
numbers=none]
unfold is_sint32 in H_sint32_i_val.
lia.
\end{lstlisting}

This closes the proof. The first repair adds the missing case split, while the second changes only the closing tactic for the boundary case. Together, continuous reasoning and localized repair allow \toolname\ to complete the lemma; the final proof is shown in Listing~\ref{lst-motivating-final}.


\begin{lstlisting}[
style=CoqStyle,
numbers=none,
caption={Final proof generated by ProofCoder.},
label={lst-motivating-final}]
intros t i i1 ... H_sint32_i H_sint32_succ H_sint32_0 H_sint32_i_val H_forall.
unfold x.
assert (i1 < i \/ i1 = i) as [H_lt | H_eq] by lia.
apply H_forall.
assumption.
assumption.
rewrite H_eq.
unfold is_sint32 in H_sint32_i_val.
lia.
\end{lstlisting}

\vskip 0.3em
\noindent\noindent
\emph{\textbf{Takeaway.}}
This example shows that the key difference is not whether the agents find the right high-level idea: both AutoRocq and \toolname\ identify the split between \texttt{i1 < i} and \texttt{i1 = i}. The difference is how they react to the resulting proof states. AutoRocq reaches the right decomposition but fails to close the boundary case, while \toolname\ uses each failed tactic to revise its local plan.

This is where focused repair matters. The failed steps do not invalidate the whole proof: the introduced context, the unfolding of \texttt{x}, the case split, and the solved \texttt{i1 < i} branch remain useful. \toolname\ can preserve this prefix and replace only the failing fragment. Its \emph{continuous reasoning between
tool calls becomes the main driver in proof repair}, 
allowing the agent to interpret proof-state feedback more accurately and fix the script, first adding the missing case split and then unfolding \texttt{is\_sint32} to close the final arithmetic goal.

\section{Agentic Proof Synthesis with Continuous Reasoning}

\begin{figure}
    \centering
    \includegraphics[width=1\linewidth]{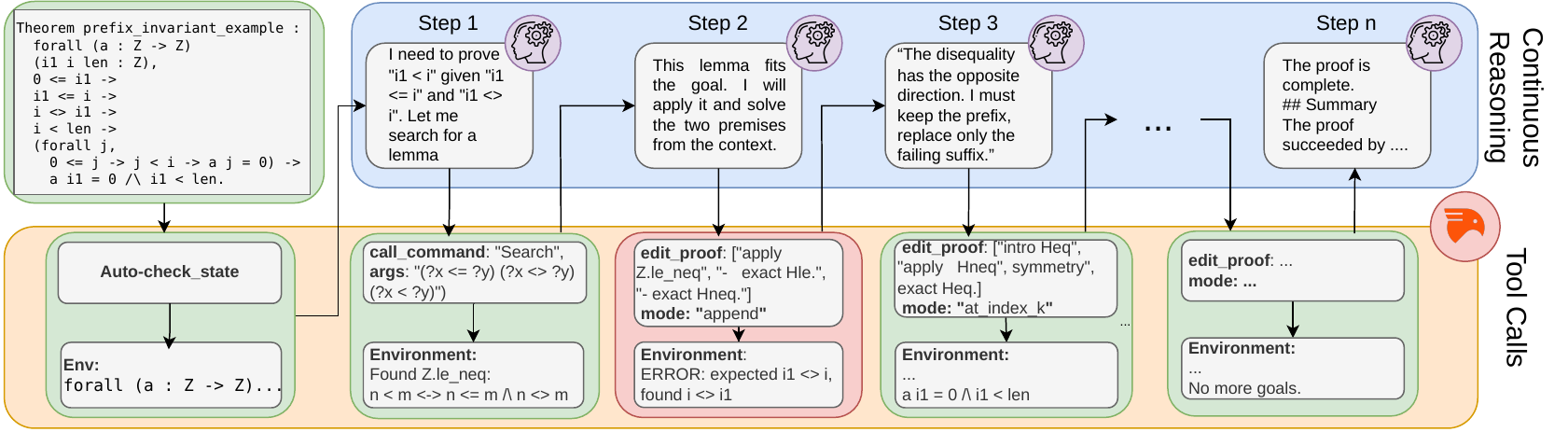}
    \caption{Overview of ProofCoder's workflow.}
    \label{fig:proofcoder_diagram}
\end{figure}

Figure~\ref{fig:proofcoder_diagram} summarizes \toolname's workflow. Inspired by coding agents, \toolname\ treats the proof script as an editable artifact: at each step, it uses Rocq feedback to reason about the current state, retrieve context when needed, and either extend the proof or revise a previously generated suffix. This gives the agent control over when and where to construct and repair proofs.

\subsection{Proof Sessions}

Each proof attempt is executed inside a preconfigured Rocq session. The harness binds the agent to a single proof obligation, including the theorem name, the source file, and the workspace containing the generated verification condition. The agent does not invoke Rocq directly or edit files outside this workspace; interactions with the proof assistant go through the tool interface described below.

At the beginning of a session, the agent receives the initial proof state, including the current hypotheses and goals. From there, it incrementally constructs a proof script. The script is represented as an ordered list of tactic lines, and the proof assistant is used as the verifier after every edit. A session terminates successfully when Rocq accepts the generated script and all goals are discharged. The agent is not allowed to use placeholder proof commands such as \texttt{admit} or \texttt{shelve}/\texttt{unshelve}; only fully checked proofs are counted as successful. Furthermore, the environment's setup prevents the agent from changing the theorem statement.

\subsection{Tool Interface}

The agent interacts with Rocq through a small set of tools. These tools expose both read-only operations, used to inspect proof states or retrieve context, and a single mutating operation, used to edit the proof script.
\vskip 0.3em
\noindent\noindent
{\texttt{check\_state}.}
Read-only tool that
replays the proof script up to a given tactic index and returns the corresponding proof state. The index denotes the length of the prefix to execute: index \texttt{0} corresponds to the initial goal, while index \texttt{n} corresponds to the state after the first \texttt{n} tactic lines. This allows the agent to inspect not only the current state, but also earlier states in the proof. In practice, the agent should use this tool sparingly, primarily when it needs to revisit the context at an earlier point before deciding where to repair the script.
\vskip 0.3em
\noindent\noindent
{\texttt{call\_command}.}
Read-only tool that
provides access to standard Rocq commands such as \texttt{Search}, \texttt{Print} and \texttt{Locate}. The agent can search for relevant lemmas, inspect definitions, locate identifiers, or check the type of a term before using it in the proof. The output of these commands is returned as text and becomes part of the subsequent model context.
\vskip 0.3em
\noindent\noindent
{\texttt{read\_proof\_file}.}
Read-only tool that enables the agent to
inspect files in the session workspace. This is useful when the compact proof-state display omits information that is easier to understand from the source theorem, or when the agent needs to verify definitions and local context present in the file.
\vskip 0.3em
\noindent\noindent
{\texttt{edit\_proof}.}
This is the only tool that modifies the proof script. The tool takes a proof fragment and inserts it into the current script. By default, the fragment is appended to the end of the proof. Alternatively, the agent may provide an edit index \texttt{n}. In that case, the environment preserves the first \texttt{n} proof lines, discards the remaining suffix, and appends the newly generated fragment.

This interface is independent of the granularity of proof generation. In a tactic-level configuration, the generated fragment is restricted to a single tactic. In a whole-proof configuration, the fragment may contain an entire proof attempt or the suffix necessary to complete it. Thus, it provides a uniform mechanism for forward proof construction, localized repair, and full proof revision.

After each edit, the environment rechecks the resulting script and returns feedback from Rocq. If the edit is accepted but goals remain, the response contains the updated proof state. On failure, the response reports the failing tactic index and the corresponding proof state at the point of failure. The agent can then continue from the current prefix, replace the rejected step, or backtrack further by editing from an earlier index.




\subsection{Continuous Reasoning}
The integration of continuous reasoning~\cite{react} with verifier-guided proof repair is a central part of our agent design. At each step, the model produces a natural language message, along with the tool call, recording the model's interpretation of the current proof state, the local proof objective, the intended action, and, when relevant, its diagnosis of a previous failed attempt. The tool call executes the action described in this reasoning, such as inspecting a proof state, querying Rocq, or editing the proof script.

The role of these messages is to guide the agent's control policy. Proof search often requires decisions that are not captured by the next tactic alone: whether a failed subgoal indicates a minor syntactic error, a wrong lemma application, an unhelpful case split, or a proof prefix that should be abandoned. By making this diagnosis explicit before each action, the agent maintains a running account of the proof attempt that conditions later decisions.

This mechanism is especially important for proof repair. When an edit fails or leads to an unproductive state, the agent can use its previous reasoning together with the new proof-state feedback to decide how much of the proof should be preserved. For example, it may conclude that only the last tactic should be replaced, that an earlier assertion was formulated incorrectly, or that the proof should restart from a previous case split. Continuous reasoning therefore supports not only tactic selection, but also the higher-level decision of where and how to revise the proof script.

This differs from systems where the model primarily emits isolated tactics, or where an external proof-search loop predetermines when to query the environment and when to backtrack, such as Rango~\cite{rango} or PALM~\cite{palm}. 
It also differs from AutoRocq, where reasoning is organized primarily around an explicit initial plan rather than a continuous rationale before each proof-editing action. In our setting, the model controls both the proof edits and the surrounding reasoning process. The resulting workflow resembles coding-agent repair: progress is made by repeatedly editing an artifact, observing the result, diagnosing the failure mode, and revising the artifact accordingly.

\section{Experimental Setup}
This section describes the experimental setup, including the research questions, benchmark, baselines, implementation and configurations, and stopping criteria.

\subsection{Research Questions}
To evaluate the effectiveness of \toolname\ on verification tasks, our evaluation answers three research questions:
\def\RQOne{How does ProofCoder compare with the state of the art, and which design choices contribute most to its performance?}
\def\RQTwo{How do different agentic configurations affect backtracking behavior during proof search?}
\def\RQThree{How do different proof generation granularity strategies compare in our setup?}
\begin{itemize}[leftmargin=*]
    \item \textbf{RQ1.} \RQOne
    \item \textbf{RQ2.} \RQTwo
    \item \textbf{RQ3.} \RQThree
\end{itemize}

\subsection{Benchmarks}

\noindent\noindent
\emph{\textbf{SV-COMP.}}
We consider the SV-COMP dataset of the AutoRocq benchmark~\cite{autorocq}, consisting of verification lemmas generated from annotated C programs through a deductive-verification workflow. In the Frama-C ecosystem~\cite{kirchner2015frama}, these programs are typically specified in ACSL~\cite{acsl}, and the resulting verification conditions can be discharged by automatic provers, for example through Why3~\cite{filliatre2013why3}, or exported to interactive proof assistants when more expressive reasoning is required.
Due to cost constraints, we select the \textbf{200 verification lemmas} with the highest term count, one of the complexity metrics used by \citet{autorocq}. Their term counts range from 178 to 703, with a mean of 292.4 and a median of 265.5. Since no ground-truth proofs are available, the benchmark does not establish which lemmas are provable; however, 88 tasks in our subset were solved by AutoRocq or one of its baselines in the original evaluation. Thus, while provability is not established for the full subset, it retains most obligations known to be solvable by existing approaches and provides a challenging setting without being dominated by obligations for which no proof has yet been found.

\noindent\noindent
\emph{\textbf{Linux Kernel.}}
We select \textbf{80 verification lemmas} from Verker, a benchmark that formalizes functional correctness of Linux kernel code, used in AutoRocq's own case study.
As reported by \citet{autorocq}, these lemmas are used to reason about the correctness of intricate program behaviors, and come from various source files, such as memory management (e.g., \textsf{memmove}) and utilities such as string operations (e.g., \textsf{strcpy}) and type conversion (e.g., \textsf{hex2bin}). In terms of complexity, these lemmas have term counts ranging from 29 to 549, with a mean of 178.66 and a median of 130.

\subsection{Baselines}

We compare \toolname\ against several existing theorem-proving systems, using AutoRocq~\cite{autorocq} as our primary baseline because it is the state of the art in agentic Rocq-based software verification and the closest system to our approach.
In its original evaluation, AutoRocq also compared against Rango~\cite{rango}, PALM~\cite{palm}, COPRA~\cite{copra}, Proverbot9001~\cite{prooverbot9001}, and QEDCartographer~\cite{qedcartographer}, outperforming these systems.
We reuse these published baseline results to provide broader context, while focusing our new experiments on AutoRocq.

AutoRocq is an agentic loop with access to four tools: \emph{Plan}, which allows it to generate a free-form plan for the proof, stored in the context; \emph{Query}, to execute query commands like \texttt{Search}, \texttt{Print} and \texttt{Locate}; \emph{Tactic}, used to generate tactics; and \emph{Rollback}, used to delete a certain number of previously accepted steps. \emph{Rollback} was not evaluated originally in \cite{autorocq}, but the tool implementation was present in the replication package and early experiments showed that, with more recent models, it improved the performance.
AutoRocq processes the proof script and proof state as a tree. Context retrieval, either by using query commands or by feeding proofs from history, is usually done at the beginning of a new state or after a certain number of failed attempts to generate a tactic.

\def\AutoRocqRollback{AutoRocq}
\def\ProofCoderWP{\toolname-WP}
\def\ProofCoderTL{\toolname-TL}
\def\ProofCoderTLRollback{\toolname-TL-NoCR+Rollback}
\def\ProofCoderTLNoCR{\toolname-TL-NoCR}

\def\sonnet{\texttt{claude-sonnet-4.5}}
\def\sol{\texttt{gpt-6-sol}}

AutoRocq was originally evaluated using \texttt{gpt-4.1}. We revisit this design choice by replacing it with the more recent models \sonnet\ and \sol.

As previously mentioned, one further challenge when evaluating theorem provers on AutoRocq  benchmark is that ground-truth proofs are generally unavailable. Consequently, some obligations may be unprovable, or simply beyond the capabilities of current automated approaches. 
To account for this, the agent terminates a proof attempt when it determines that further search is unlikely to succeed. We refer to this behavior as \emph{give up}. 
In order to save costs and ensure a fair comparison across approaches, we equip our agent with an equivalent mechanism in our evaluation.

\subsection{Agent Implementation and Configurations}

We implement \toolname\ as a thin integration layer on top of the OpenCode agent runtime~\cite{opencode}, connected to Rocq through a custom Model Context Protocol (MCP) server of approximately 3,100 lines of Python. This separates agent orchestration, including reasoning, tool selection, and conversation management, from theorem-prover interaction, including script maintenance, replay, and goal reporting. The Rocq backend reuses AutoRocq's CoqInterface, based on CoqPyt~\cite{CoqPyt}, coq-lsp~\cite{rocq-lsp}, and Rocq 8.18, which the MCP server imports at runtime rather than reimplementing. We run our experiments across  \sonnet\ and \sol\, both with temperature 0.

We define the following \toolname\ configurations, where WP and TL stand for, respectively, Whole Proof and Tactic-Level:
\begin{itemize}
    \item \emph{\ProofCoderWP}: 
    the agent generates multiple tactics per \emph{edit\_proof} call and is prompted to always try to complete the proof. It repairs proofs through the default behavior of \emph{edit\_proof}.
    \item \emph{\ProofCoderTL}: 
    the agent generates one tactic per \emph{edit\_proof} call. It repairs proofs through the default behavior of \emph{edit\_proof}.
    \item \emph{\ProofCoderTLNoCR}: an ablation that uses a plan tool instead of continuous reasoning between tool calls (tactic-level).
    \item \emph{\ProofCoderTLRollback}: an ablation that more closely matches AutoRocq: the agent performs tactic-level generation, backtracks only through \emph{Rollback} and produces no continuous reasoning between calls.
    \item \emph{ProofCoder-WP-NoCR}: uses a plan tool instead of continuous reasoning between tool calls (whole proof).
    \item \emph{ProofCoder-WP-Reset}: replaces localized backtrack with a deterministic reset that restarts the proof after every failed generation.
    
\end{itemize}

\subsection{Stopping Criteria}
Evaluating agentic provers can be computationally expensive, so we impose a fixed resource budget on each proof attempt. Limits were set on the basis of practical cost constraints and preliminary experiments, which showed diminishing improvements in proof success beyond these thresholds.

Our primary stopping criterion is a budget of 200k non-cached tokens. This metric is better suited to comparing agents with different proof-generation granularities than the number of tool calls, which would favor some configurations over others. Non-cached token usage provides a more consistent measure of the model usage required for a proof attempt.

We impose two secondary limits for edge cases. First, we cap total token usage at 5M tokens, since some tactic-level configurations can produce very long interactions while consuming relatively few non-cached tokens due to context reuse. Second, we set a 30-minute wall-clock timeout per verification lemma to prevent a small number of unusually expensive searches from dominating the evaluation time. These additional limits make the experiments predictable while having little impact on typical proof attempts.

\section{Experimental Results}
This section presents the experimental results and answers our research questions.

\subsection{RQ1. \RQOne}

Table~\ref{tab:svcomp-comparison} reports proof success on the SV-COMP lemmas, including our \toolname\ configurations, AutoRocq using the same models (\sonnet\ and \sol), and the baselines reported in the original AutoRocq evaluation. AutoRocq benefits substantially from the stronger model, improving from 86 to 94 solved lemmas when moving from \sonnet\ to \sol, which shows that its agentic design continues to benefit from advances in the underlying LLM. \toolname\ shows the same trend and, importantly, both of its configurations outperform AutoRocq with either model. With \sonnet, \ProofCoderWP\ proves 113 lemmas, 27 more than \AutoRocqRollback, a 31.4\% relative improvement, while \ProofCoderTL\ proves 25 more. With \sol, \ProofCoderWP\ and \ProofCoderTL\ prove 26 and 25 more lemmas than \AutoRocqRollback, respectively.

\begin{figure}[htbp]
\resizebox{0.8\linewidth}{!}{%
\begin{minipage}{\linewidth}
    \centering
    \begin{subfigure}[b]{0.24\linewidth}
        \centering
        \includegraphics[width=0.9\linewidth]{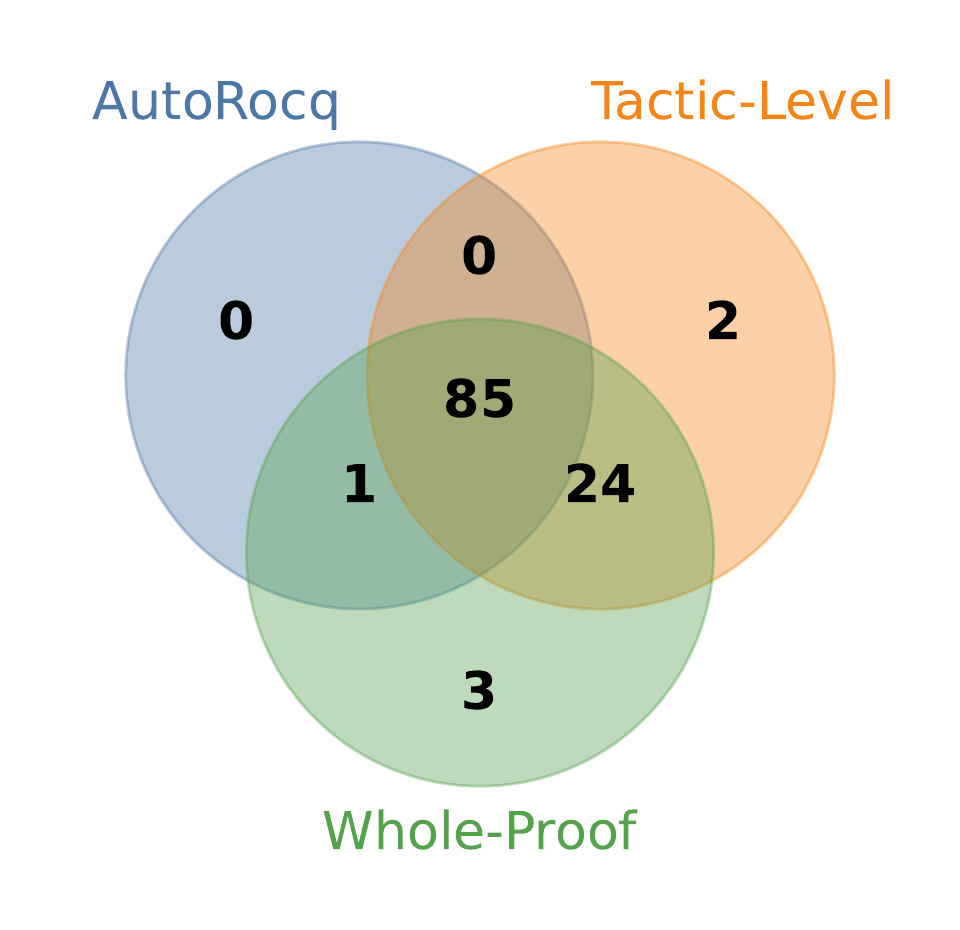}
        \caption{SV-COMP (Sonnet 4.5)}
        \label{fig:svcomp_venn_sonnet}
    \end{subfigure}
    \hfill
    \begin{subfigure}[b]{0.24\linewidth}
        \centering
        \includegraphics[width=0.9\linewidth]{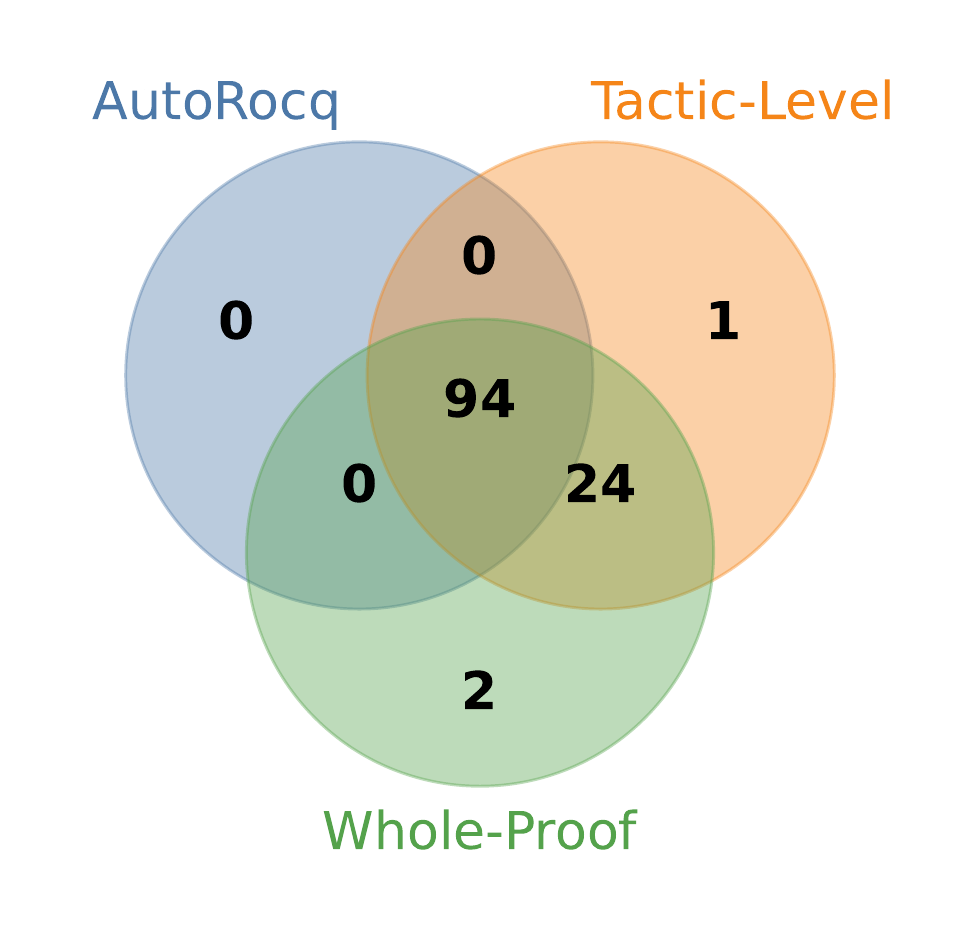}
        \caption{SV-COMP (gpt-6-sol)}
        \label{fig:svcomp_venn_gpt}
    \end{subfigure}
    \hfill
    \begin{subfigure}[b]{0.24\linewidth}
        \centering
        \includegraphics[width=0.9\linewidth]{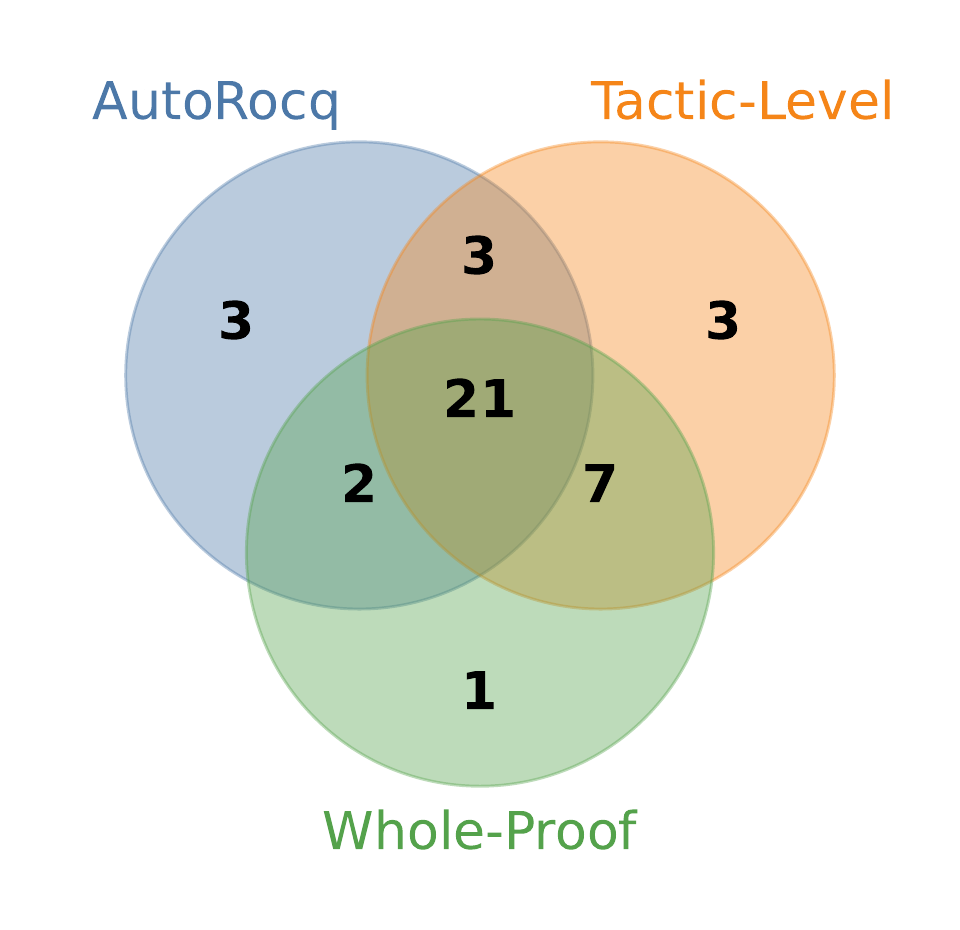}
        \caption{Verker (Sonnet 4.5)}
        \label{fig:verker_venn_sonnet}
    \end{subfigure}
    \hfill
    \begin{subfigure}[b]{0.24\linewidth}
        \centering
        \includegraphics[width=0.9\linewidth]{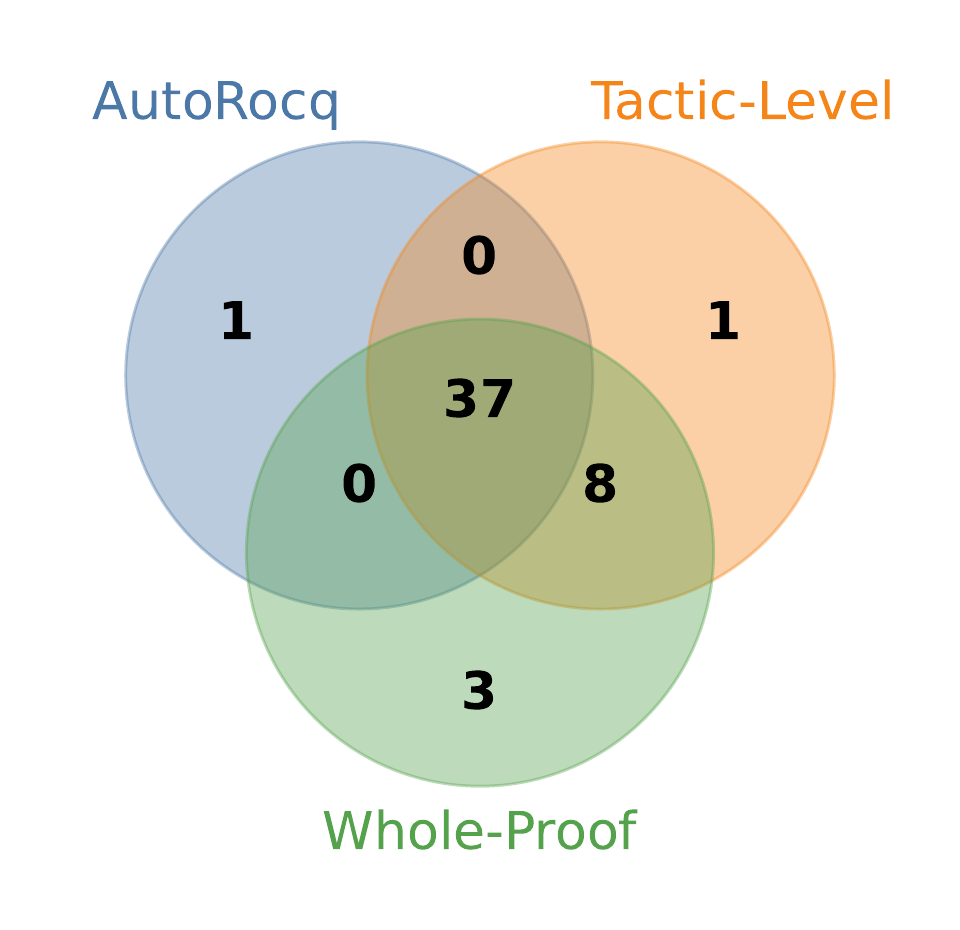}
        \caption{Verker (gpt-6-sol)}
        \label{fig:verker_venn_gpt}
    \end{subfigure}
\end{minipage}
}
    \caption{Overlap of verification lemmas proved by AutoRocq and ProofCoder.}
    \label{fig:svcomp_venn_diagram}
\end{figure}

Furthermore, as shown in Figures~\ref{fig:svcomp_venn_sonnet} and \ref{fig:svcomp_venn_gpt}, all verification lemmas proved by AutoRocq were also proved by at least one configuration of \toolname. 

Because the agentic systems substantially outperform the other baselines on SV-COMP, we compare our approaches exclusively with AutoRocq on Verker. In Table~\ref{tab:verker-comparison}, we see a similar difference in performance between approaches, with \ProofCoderWP\ proving 38.8\% and 60\% of lemmas with \sonnet\ and \sol, respectively and \ProofCoderTL\ proving 42.5\% and 57.5\%. AutoRocq, on the other hand, proved 36.2\% and 47.5\%.

The Venn diagrams in Figures~\ref{fig:verker_venn_sonnet} and~\ref{fig:verker_venn_gpt} show that, although our approach outperforms \AutoRocqRollback, the two approaches do not solve exactly the same lemmas: a small number are proved exclusively by \AutoRocqRollback. 
This suggests that, on the more challenging Verker dataset, different agentic approaches can still succeed on different proof obligations, with our best configuration proving three AutoRocq-unsolved lemmas compared with one lemma solved exclusively by \AutoRocqRollback. 

\begin{table}[t]
\caption{Pass rates, average cost, and average time on 200 SV-COMP lemmas. The last three columns are solved lemmas of each obligation type (49 assert reachability, 30 signed int overflow, and 121 loop invariant).}
\centering
\small
\resizebox{\linewidth}{!}{%
\begin{tabular}{lrrrrrr}
\toprule
\textbf{System} & \textbf{Passed} & \textbf{Avg.\ cost (USD)} & \textbf{Avg.\ time (s)} &
\begin{tabular}[b]{@{}c@{}}\textbf{Assert}\\\textbf{reachability}\end{tabular} &
\begin{tabular}[b]{@{}c@{}}\textbf{Signed int}\\\textbf{overflow}\end{tabular} &
\begin{tabular}[b]{@{}c@{}}\textbf{Loop}\\\textbf{invariant}\end{tabular} \\
\midrule
Rango & 40 (20.0\%) & --- & --- & 22 (44.9\%) & 4 (13.3\%) & 14 (11.6\%) \\
PALM & 24 (12.0\%) & --- & --- & 6 (12.2\%) & 1 (3.3\%) & 17 (14.0\%) \\
QEDCartographer & 46 (23.0\%) & --- & --- & 25 (51.0\%) & 2 (6.7\%) & 19 (15.7\%) \\
Proverbot9001 & 56 (28.0\%) & --- & --- & 24 (49.0\%) & 2 (6.7\%) & 30 (24.8\%) \\
COPRA & 71 (35.5\%) & --- & --- & 23 (46.9\%) & 8 (26.7\%) & 40 (33.1\%) \\
AutoRocq (GPT-4.1) & 70 (35.0\%) & --- & --- & 21 (42.9\%) & 11 (36.7\%) & 38 (31.4\%) \\
\midrule
AutoRocq (\sonnet) & 86 (43.0\%) & 0.678 & \textbf{609} & 30 (61.2\%) & 22 (73.3\%) & 34 (28.1\%) \\
Tactic-Level (\sonnet) & 111 (55.5\%) & 0.516 & 747 & 31 (63.3\%) & 22 (73.3\%) & 58 (47.9\%) \\
Whole-Proof (\sonnet) & \textbf{113 (56.5\%)} & \textbf{0.352} & 895 & \textbf{32 (65.3\%)} & \textbf{23 (76.7\%)} & \textbf{58 (47.9\%)} \\
\midrule
AutoRocq (\sol) & 94 (47.0\%) & 0.177 & 475 & 29 (59.2\%) & 20 (66.7\%) & 45 (37.2\%) \\
Tactic-Level (\sol) & 119 (59.5\%) & 0.090 & 230 & 33 (67.3\%) & \textbf{25 (83.3\%)} & \textbf{61 (50.4\%)} \\
Whole-Proof (\sol) & \textbf{120 (60.0\%)} & \textbf{0.084} & \textbf{229} & \textbf{34 (69.4\%)} & \textbf{25 (83.3\%)} & \textbf{61 (50.4\%)} \\
\bottomrule
\end{tabular}}
\label{tab:svcomp-comparison}
\end{table}

\begin{table}[t]
\caption{Pass rates, average cost, and average time on 80 Verker lemmas. The last two columns are solved lemmas of each obligation type (53 assert and 27 loop invariant).}
\centering
\small
\resizebox{0.88\linewidth}{!}{%
\begin{tabular}{lrrrrr}
\toprule
\textbf{System} & \textbf{Passed} & \textbf{Avg.\ cost (USD)} & \textbf{Avg.\ time (s)} &
\begin{tabular}[b]{@{}c@{}}\textbf{Assert}\\\textbf{reachability}\end{tabular} &
\begin{tabular}[b]{@{}c@{}}\textbf{Loop}\\\textbf{invariant}\end{tabular} \\
\midrule
AutoRocq (\sonnet) & 29 (36.2\%) & 1.336 & 856 & 17 (32.1\%) & \textbf{12 (44.4\%)} \\
Tactic-Level (\sonnet) & \textbf{34 (42.5\%)} & 1.177 & \textbf{727} & \textbf{23 (43.4\%)} & 11 (40.7\%) \\
Whole-Proof (\sonnet) & 31 (38.8\%) & \textbf{0.970} & 1007 & 22 (41.5\%) & 9 (33.3\%) \\
\midrule
AutoRocq (\sol) & 38 (47.5\%) & 0.388 & 653 & 27 (50.9\%) & 11 (40.7\%) \\
Tactic-Level (\sol) & 46 (57.5\%) & 0.288 & \textbf{391} & 31 (58.5\%) & \textbf{15 (55.6\%)} \\
Whole-Proof (\sol) & \textbf{48 (60.0\%)} & \textbf{0.182} & 420 & \textbf{33 (62.3\%)} & \textbf{15 (55.6\%)} \\
\bottomrule
\end{tabular}}
\label{tab:verker-comparison}
\end{table}

\emph{Cost and Time Analysis:}
Whole-Proof is the cheapest of the three systems, and AutoRocq is the most expensive. On SV-COMP with \sonnet, \ProofCoderWP\ (\$0.352) costs about two-thirds as much as \ProofCoderTL\ (\$0.516) and about half as much as AutoRocq (\$0.678). With \sol\ the two agents are close: \$0.084 against \$0.090, both about half of AutoRocq (\$0.177). On Verker, the Whole-Proof saving is larger for \sol\ (\$0.182 against \$0.288 and \$0.388) than for \sonnet\ (\$0.970 against \$1.177 and \$1.336).

Time does not follow cost. With \sonnet, \ProofCoderWP\ is the slowest configuration on both datasets (895 s on SV-COMP, 1007 s on Verker), while \ProofCoderTL\ is faster than Whole-Proof and, on Verker, faster than AutoRocq (727 s against 856 s). With \sol\ the two agents take almost the same time, and both are faster than AutoRocq: about 230 s against 475 s on SV-COMP, and about 391–420 s against 653 s on Verker. Verker is the more expensive dataset for every system, at roughly two to three times the SV-COMP cost per lemma, reflecting the increase in difficulty for this dataset.

\emph{Verification Lemma Categories:}
The SV-COMP tasks fall into three categories: Assert Reachability (49 lemmas), Assert (RTE) Signed Overflow (30 lemmas), and Loop Invariants (121 lemmas). As shown in Table~\ref{tab:svcomp-comparison}, \toolname's gains over AutoRocq are concentrated on loop invariants. With \sonnet, AutoRocq solves 34 of these lemmas (28.1\%), while \ProofCoderTL\ and \ProofCoderWP\ each solve 58 (47.9\%). That increase of 24 lemmas is the whole net gain of Tactic-Level over AutoRocq (111 against 86) and 24 of Whole-Proof's 27 additional proofs (113 against 86). On the other two categories \sonnet\ AutoRocq is already close to the agents: 30 against 31 and 32 of the 49 assert-reachability lemmas, and 22 against 22 and 23 of the 30 signed-overflow lemmas. With \sol\ the same concentration holds, above a stronger AutoRocq baseline. AutoRocq solves 45 loop invariants (37.2\%), and both agents solve 61 (50.4\%), which is 16 of the 25 and 26 extra proofs by \ProofCoderTL\ and \ProofCoderWP, respectively. The rest is split across the other categories: assert reachability rises from 29 to 33 and 34, and signed overflow from 20 to 25 for both agents.

Loop invariant lemmas are not substantially more complex by syntactic metrics, such as hypothesis count, term count, or statement length. They differ in proof structure, requiring reasoning about multiple memory snapshots, universally quantified array properties, map updates, and case distinctions over indices. This makes them a natural setting for agents that inspect intermediate proof states, reason continuously about the evolving context, and revise partially constructed proofs. Manual inspection suggests that \toolname's advantage is not primarily due to discovering entirely new proof strategies. In more than half of the tasks solved only by \toolname, AutoRocq attempted the same key proof steps, such as unfolding \texttt{Map.set}, splitting on index equalities, or instantiating universally quantified hypotheses. The difference lies in execution: \toolname\ preserves promising prefixes, inspects previous states, and repairs localized failures, while AutoRocq often remains stuck in local repairs until eventually resetting the proof through rollback.

The Verker tasks fall into two categories, Assert Reachability (53 lemmas) and Loop Invariant (27), as shown in Table~\ref{tab:verker-comparison}. Unlike SV-COMP, where \toolname's gains are concentrated on Loop Invariant lemmas, the Verker results with \sonnet\ are driven by Assert Reachability. AutoRocq solves 17 of 53 Assert Reachability lemmas (32.1\%) and 12 of 27 Loop Invariant lemmas (44.4\%), compared with 23 (43.4\%) and 11 (40.7\%) for \ProofCoderTL\ and 22 (41.5\%) and 9 (33.3\%) for \ProofCoderWP, respectively. Thus, the additional proofs obtained with \sonnet\ come entirely from Assert Reachability, while both configurations solve fewer Loop Invariant lemmas than AutoRocq. With \sol, however, the pattern changes: all three approaches improve on both categories, and both \toolname\ configurations surpass AutoRocq on Loop Invariants. AutoRocq solves 27 Assert Reachability lemmas (50.9\%) and 11 Loop Invariant lemmas (40.7\%), compared with 31 (58.5\%) and 15 (55.6\%) for \ProofCoderTL\ and 33 (62.3\%) and 15 (55.6\%) for \ProofCoderWP. This suggests that the relative advantage of \toolname\ depends not only on the proof category but also on the underlying model: the stronger model enables gains on Loop Invariants that are not observed with \sonnet\ on this dataset.
\vskip 0.3em
\noindent\noindent
\emph{{\textbf{Ablations.}}}
To better understand the impact of \toolname’s design choices, we evaluate the four ablation configurations described above using our best-performing model, \sol. Table~\ref{tab:gpt6sol-ablations} summarizes the results.
Removing continuous reasoning reduces performance while substantially increasing cost and runtime for \sol.
For \ProofCoderTL, performance drops from 119 to 115 solved lemmas on SV-COMP and from 46 to 40 on Verker, while cost more than doubles on both datasets and Verker runtime nearly doubles. \ProofCoderWP\ shows the same trend, dropping from 120 to 113 solved lemmas on SV-COMP and from 48 to 43 on Verker, with cost increasing by 55\% and 79\%, respectively. Replacing our backtrack with an explicit rollback tool to reasoning-free \ProofCoderTL\ has little effect on SV-COMP performance, which remains at 115 solved lemmas, while performance on Verker decreases to 35. These results are consistent with the hypothesis that the reasoning step helps the agent coordinate tool use and proof search.
Replacing \ProofCoderWP's localized repair with a full restart from the beginning on error leads to a much larger performance drop, from 120 to 81 solved SV-COMP lemmas and from 48 to 24 on Verker. The restart strategy also slightly increases SV-COMP cost from \$0.084 to \$0.192. This result is consistent with the hypothesis that preserving the existing proof and repairing the failing suffix is more effective than discarding the proof and restarting.

Overall, the ablation results show that continuous reasoning and localized repair are key to \toolname's effectiveness: removing either reduces proof success, with localized repair having the larger impact.
 
\begin{table}[t]
\caption{ProofCoder ablations with \sol: proof success, cost, and time on SV-COMP and Verker.}
\centering
\small
\resizebox{0.85\linewidth}{!}{%
\begin{tabular}{l|rrr|rrr}
\toprule
& \multicolumn{3}{c|}{\textbf{SV-COMP (200 lemmas)}} & \multicolumn{3}{c}{\textbf{Verker (80 lemmas)}} \\
\textbf{Ablation} & \textbf{Passed} & \textbf{Avg. cost (USD)} & \textbf{Avg. time (s)}
         & \textbf{Passed} & \textbf{Avg. cost (USD)} & \textbf{Avg. time (s)} \\
\midrule
TL-NoCR          & 115 (57.5\%) & 0.206 & 358
                 & 40 (50.0\%)  & 0.674 & 768 \\
TL-NoCR+Rollback & 115 (57.5\%) & 0.154 & 339
                 & 35 (43.8\%)  & 0.689 & 646 \\
WP-NoCR          & 113 (56.5\%) & 0.130 & 301
                 & 43 (53.8\%)  & 0.326 & 653 \\
WP-Reset         & 81 (40.5\%)  & 0.192 & 346
                 & 24 (30.0\%)  & 0.403 & 542 \\
\bottomrule
\end{tabular}}
\label{tab:gpt6sol-ablations}
\end{table}

\RQAnswer{Answer to RQ1:}
{\toolname\ consistently outperforms the state of the art across both benchmarks and both models. On SV-COMP, every lemma proved by AutoRocq is also proved by at least one \toolname\ configuration. \ProofCoderWP\ improves over AutoRocq by 31.4\% with \sonnet\ and 27.7\% with \sol, while requiring roughly half the inference cost, with gains concentrated on loop-invariant lemmas. Whole-proof and tactic-level generation achieve similar success rates, with \ProofCoderWP\ generally offering the best success--cost trade-off. The ablations show that continuous reasoning and localized repair are key design choices: removing continuous reasoning lowers success while substantially increasing cost, whereas replacing localized repair with full restarts causes much larger performance drops. Replacing our backtracking mechanism with AutoRocq's rollback tool does not improve performance.}

\subsection{RQ2. \RQTwo}


Backtracking lets agents revise accepted proof fragments rather than follow a monotonic tactic sequence, ranging from local repairs to broader exploration or full restarts. Its behavior depends on the interaction interface, including rollback mechanisms, editable proof scripts, and continuous reasoning. Effective backtracking therefore requires deciding which proof prefix to preserve and which suffix to discard. We compare backtracking across tactic-level approaches by analyzing backtrack depth and frequency, tactic replacement patterns, and the fraction of preserved revisions. We use the \sol configuration of our agent and its ablations.

\vskip 0.3em
\noindent\noindent
\emph{\textbf{Backtrack depth and frequency.}}
We define \emph{backtrack depth} as the percentage of the current proof discarded by a backtrack, and \emph{backtrack frequency} as the mean number of backtracks per task. Depth captures repair aggressiveness: higher values indicate larger discarded suffixes, while lower values indicate local edits near the end of the script.

\begin{figure}[t]
\centering
\subfloat[Backtrack depth and frequency (SV-COMP)\label{fig:svcomp_backtrack_depth_frequency}]{%
  \includegraphics[width=0.49\linewidth]{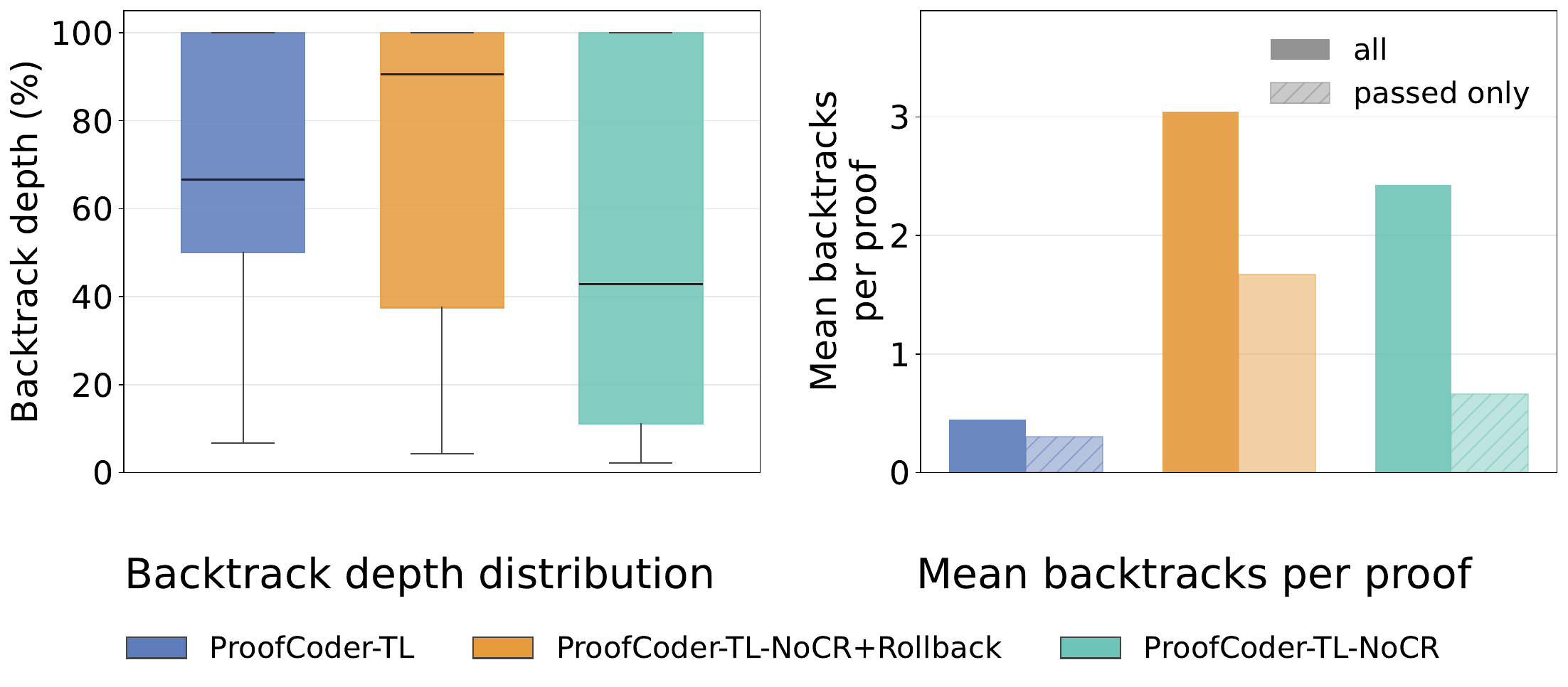}%
}\hfill%
\subfloat[Backtrack depth and frequency (Verker)\label{fig:verker_backtrack_depth_frequency}]{%
  \includegraphics[width=0.49\linewidth]{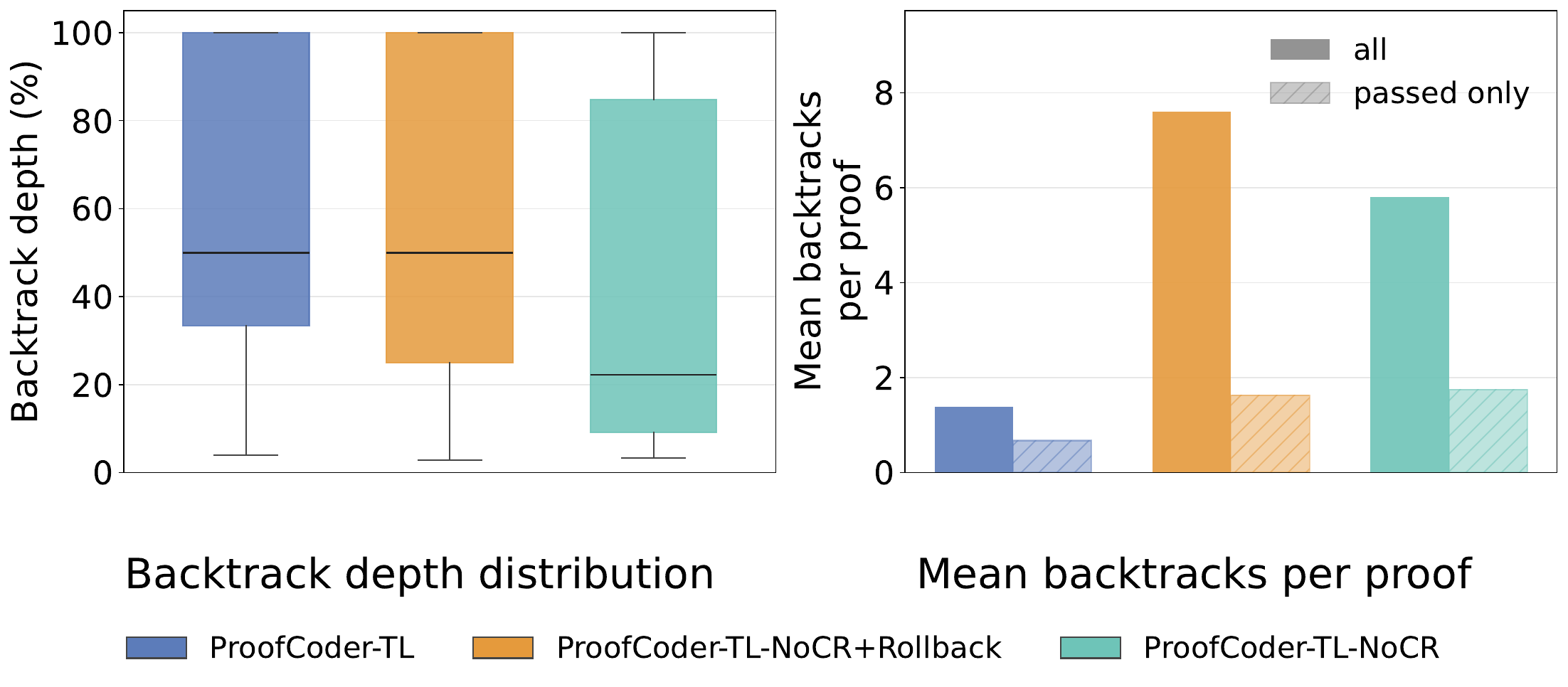}%
}
\caption{Comparison of backtrack depth and frequency across tactic-level approaches.}
\label{fig:backtrack_depth_frequency}
\end{figure}

Figure~\ref{fig:backtrack_depth_frequency} shows substantial differences between approaches. The most pronounced effect of continuous reasoning is on frequency: \ProofCoderTL\ backtracks roughly 4--6$\times$ less often than both ablations (0.45 vs.\ 2.4 for \ProofCoderTLNoCR\ and 3.05 for \ProofCoderTLRollback\ on SV-COMP; 1.39 vs.\ 5.8 and 7.6 on Verker). Moreover, for the ablations, the frequency over all tasks is 2--5$\times$ as high than over successful tasks alone, indicating that heavy backtracking is characteristic of failing runs, whereas \ProofCoderTL\ shows only a small gap between the two. The approaches also differ in how they backtrack. Without reasoning, \ProofCoderTLNoCR\ favors shallow, local edits (median depth $\approx$43\% on SV-COMP and $\approx$22\% on Verker), while \ProofCoderTLRollback\ tends towards near-complete restarts on SV-COMP (median depth $\approx$90\%) and, even in successful proofs, backtracks the most of the three variants. ProofCoder-TL sits between these extremes (median depth $\approx$67\% on SV-COMP and $\approx$50\% on Verker). This suggests that reasoning about a failure before acting lets the agent choose a single, appropriately placed resumption point, instead of repeatedly patching locally or discarding most of the proof.

\vskip 0.3em
\noindent\noindent
\emph{\textbf{Tactic replacement patterns.}} 
Tables~\ref{tab:svcomp-replacements} and~\ref{tab:verker-replacements} report the five most frequently deleted tactics and their three most common replacements.
Across tactic-level variants, \texttt{intros} is by far the most frequently deleted tactic. For \ProofCoderTL, it is replaced by \texttt{intros} in 92.3\% of cases on SV-COMP and 89.7\% on Verker. Removing continuous reasoning changes this substantially: \ProofCoderTLNoCR\ retains \texttt{intros} in only 34.5\% and 31.7\% of cases, respectively, with replacements dispersed across tactics such as \texttt{lia}, \texttt{pose}, and \texttt{idtac}. This suggests that, without reasoning, the chosen resumption point does not always correspond to the proof decision the agent intends to revise.

\ProofCoderTLRollback\ shows a different pattern. It deletes \texttt{intros} much more often than either other variant (337 and 239 times on SV-COMP and Verker, compared with 142 and 101 for \ProofCoderTLNoCR\ and 52 and 68 for \ProofCoderTL) and replaces it with \texttt{intros} or \texttt{intro} in over 95\% of cases. This is consistent with rollback frequently discarding the proof setup, which the agent then reconstructs before diverging later. Beyond setup, \ProofCoderTLRollback\ resembles \ProofCoderTLNoCR, with dispersed revisions among tactics such as \texttt{assert}, \texttt{destruct}, \texttt{pose}, and \texttt{exfalso}.

Other frequently revised tactics include \texttt{apply}, \texttt{assert}, and \texttt{destruct}, which correspond to substantive proof decisions such as lemma selection, intermediate facts, and case decomposition. They are often replaced by the same or closely related tactics (e.g., \texttt{apply} by \texttt{apply} in 54.3\% of cases for \ProofCoderTLNoCR\ on Verker and \texttt{assert} by \texttt{assert} in 48.6\% for \ProofCoderTLRollback), suggesting that the agent often preserves the overall strategy while revising its realization. For \ProofCoderTL, however, deletions beyond \texttt{intros} are rare ($n \leq 9$), so these replacement distributions should be interpreted cautiously.

Overall, continuous reasoning primarily reduces how often revisions are needed: \ProofCoderTL\ performs far fewer deletions and largely preserves proof setup, whereas the reasoning-free variants revise more aggressively. Rollback stabilizes the setup only because it is repeatedly reconstructed after near-complete restarts.

\begin{table}[t]
\caption{Most frequently deleted tactics and their three most common replacements on SV-COMP.}
\centering
\fontsize{6.8pt}{7.4pt}\selectfont
\setlength{\tabcolsep}{2pt}
\begin{tabularx}{\linewidth}{@{}l r L L L@{}}
\toprule
\multicolumn{5}{c}{\textbf{\ProofCoderTL}} \\
\toprule
Deleted tactic & $n$ & \#1 & \#2 & \#3 \\
\midrule
\texttt{intros}     & 52 & \rep{intros}{92.3} & \rep{cbn}{5.8} & \rep{idtac}{1.9} \\
\texttt{apply}      & 6  & \rep{apply}{33.3} & \rep{destruct}{33.3} & \rep{rewrite}{16.7} \\
\texttt{eauto}      & 6  & \rep{congruence}{33.3} & \rep{exfalso}{33.3} & \rep{nia}{16.7} \\
\texttt{destruct}   & 4  & \rep{specialize}{50.0} & \rep{assert}{25.0} & \rep{destruct}{25.0} \\
\texttt{specialize} & 3  & \rep{destruct}{66.7} & \rep{specialize}{33.3} & \\
\toprule
\multicolumn{5}{c}{\textbf{\ProofCoderTLRollback}} \\
\toprule
Deleted tactic & $n$ & \#1 & \#2 & \#3 \\
\midrule
\texttt{intros}   & 337 & \rep{intros}{89.9} & \rep{intro}{5.6} & \rep{cbv}{1.8} \\
\texttt{intro}    & 22  & \rep{intros}{50.0} & \rep{intro}{13.6} & \rep{cbn}{9.1} \\
\texttt{destruct} & 21  & \rep{assert}{33.3} & \rep{destruct}{28.6} & \rep{exfalso}{14.3} \\
\texttt{assert}   & 16  & \rep{assert}{37.5} & \rep{destruct}{37.5} & \rep{exfalso}{6.2} \\
\texttt{exfalso}  & 11  & \rep{pose}{18.2} & \rep{all}{9.1} & \rep{assert}{9.1} \\
\toprule
\multicolumn{5}{c}{\textbf{\ProofCoderTLNoCR}} \\
\toprule
Deleted tactic & $n$ & \#1 & \#2 & \#3 \\
\midrule
\texttt{intros}   & 142 & \rep{intros}{34.5} & \rep{lia}{10.6} & \rep{pose}{8.5} \\
\texttt{assert}   & 61  & \rep{assert}{29.5} & \rep{pose}{14.8} & \rep{destruct}{13.1} \\
\texttt{destruct} & 38  & \rep{assert}{23.7} & \rep{destruct}{21.1} & \rep{exfalso}{13.2} \\
\texttt{pose}     & 37  & \rep{assert}{27.0} & \rep{clear}{18.9} & \rep{pose}{13.5} \\
\texttt{idtac}    & 32  & \rep{idtac}{21.9} & \rep{clear}{12.5} & \rep{intros}{12.5} \\
\bottomrule
\end{tabularx}
\label{tab:svcomp-replacements}
\end{table}

\begin{table}[t]
\caption{Most frequently deleted tactics and their three most common replacements on Verker.}
\centering
\fontsize{6.8pt}{7.4pt}\selectfont
\setlength{\tabcolsep}{2pt}
\begin{tabularx}{\linewidth}{@{}l r L L L@{}}
\toprule
\multicolumn{5}{c}{\textbf{\ProofCoderTL}} \\
\toprule
Deleted tactic & $n$ & \#1 & \#2 & \#3 \\
\midrule
\texttt{intros}   & 68 & \rep{intros}{89.7} & \rep{cbn}{2.9} & \rep{cbv}{2.9} \\
\texttt{apply}    & 9  & \rep{apply}{22.2} & \rep{pose}{22.2} & \rep{eapply}{11.1} \\
\texttt{eapply}   & 4  & \rep{apply}{25.0} & \rep{assert}{25.0} & \rep{destruct}{25.0} \\
\texttt{assert}   & 3  & \rep{assert}{66.7} & \rep{destruct}{33.3} & \\
\texttt{replace}  & 3  & \rep{eapply}{33.3} & \rep{rewrite}{33.3} & \rep{unfold}{33.3} \\
\toprule
\multicolumn{5}{c}{\textbf{\ProofCoderTLRollback}} \\
\toprule
Deleted tactic & $n$ & \#1 & \#2 & \#3 \\
\midrule
\texttt{intros}   & 239 & \rep{intros}{92.5} & \rep{intro}{3.8} & \rep{cbv}{2.1} \\
\texttt{assert}   & 74  & \rep{assert}{48.6} & \rep{pose}{20.3} & \rep{destruct}{14.9} \\
\texttt{pose}     & 41  & \rep{pose}{36.6} & \rep{destruct}{26.8} & \rep{assert}{17.1} \\
\texttt{destruct} & 36  & \rep{destruct}{30.6} & \rep{assert}{27.8} & \rep{pose}{25.0} \\
\texttt{intro}    & 17  & \rep{intro}{64.7} & \rep{intros}{17.6} & \rep{pose}{17.6} \\
\toprule
\multicolumn{5}{c}{\textbf{\ProofCoderTLNoCR}} \\
\toprule
Deleted tactic & $n$ & \#1 & \#2 & \#3 \\
\midrule
\texttt{intros}   & 101 & \rep{intros}{31.7} & \rep{idtac}{17.8} & \rep{intro}{6.9} \\
\texttt{assert}   & 68  & \rep{assert}{22.1} & \rep{apply}{13.2} & \rep{destruct}{13.2} \\
\texttt{destruct} & 50  & \rep{destruct}{36.0} & \rep{assert}{16.0} & \rep{apply}{12.0} \\
\texttt{pose}     & 47  & \rep{clear}{27.7} & \rep{pose}{14.9} & \rep{apply}{12.8} \\
\texttt{apply}    & 35  & \rep{apply}{54.3} & \rep{exact}{17.1} & \rep{destruct}{8.6} \\
\bottomrule
\end{tabularx}
\label{tab:verker-replacements}
\end{table}

\vskip 0.3em
\noindent\noindent
\emph{\textbf{Proof revision stability.}}
We measure whether backtracks lead to stable proof revisions. A backtrack is \emph{preserved} if the proof search following it is not entirely undone at the end of the run. If a backtrack resumes generation from index $i$, it is preserved unless a later backtrack resumes from some index $j \leq i$, thereby discarding the branch introduced after the original backtrack.

We compute preservation at the proof level: for each successful proof with at least one backtrack, we calculate the fraction of preserved backtracks, and then average these fractions across proofs:
\[
\text{PreservationRate} = \frac{1}{|\mathcal{P}_{bt}|} \sum_{p \in \mathcal{P}_{bt}} \frac{\mathrm{backtracks\_preserved}(p)}{\mathrm{backtracks\_total}(p)} \times 100, 
\]
with $\mathcal{P}_{bt}$ as the set of successful proofs with at least one backtrack. This is a retrospective measure of stability: discarded backtracks still have diagnostic value, but the metric captures whether a repair became part of the surviving search trajectory.

\begin{figure}[t]
\centering
\subfloat[Preservation rate and distribution (SV-COMP).\label{fig:mean_perservation_rate}]{%
  \includegraphics[width=0.49\linewidth]{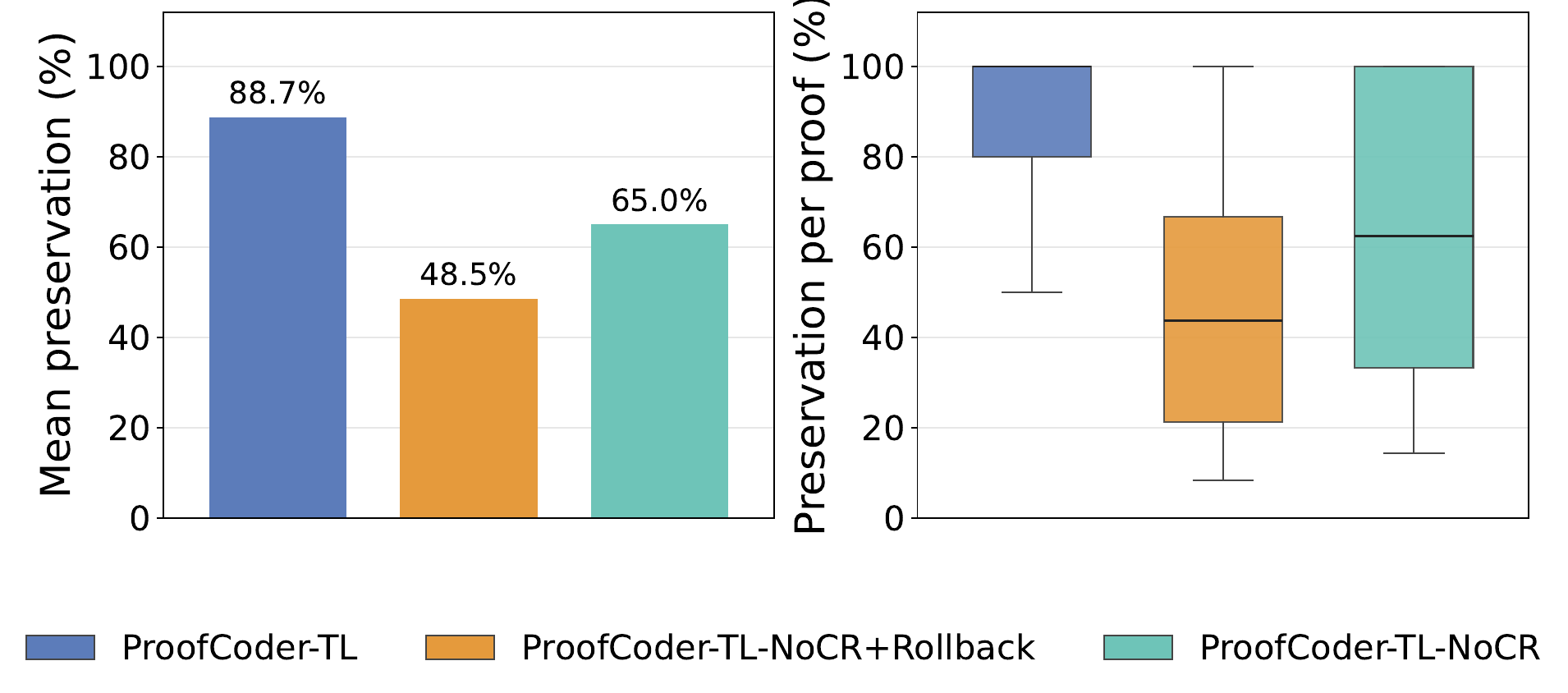}%
}\hfill%
\subfloat[Preservation rate and distribution (Verker)\label{fig:distribution_per_proof}]{%
  \includegraphics[width=0.49\linewidth]{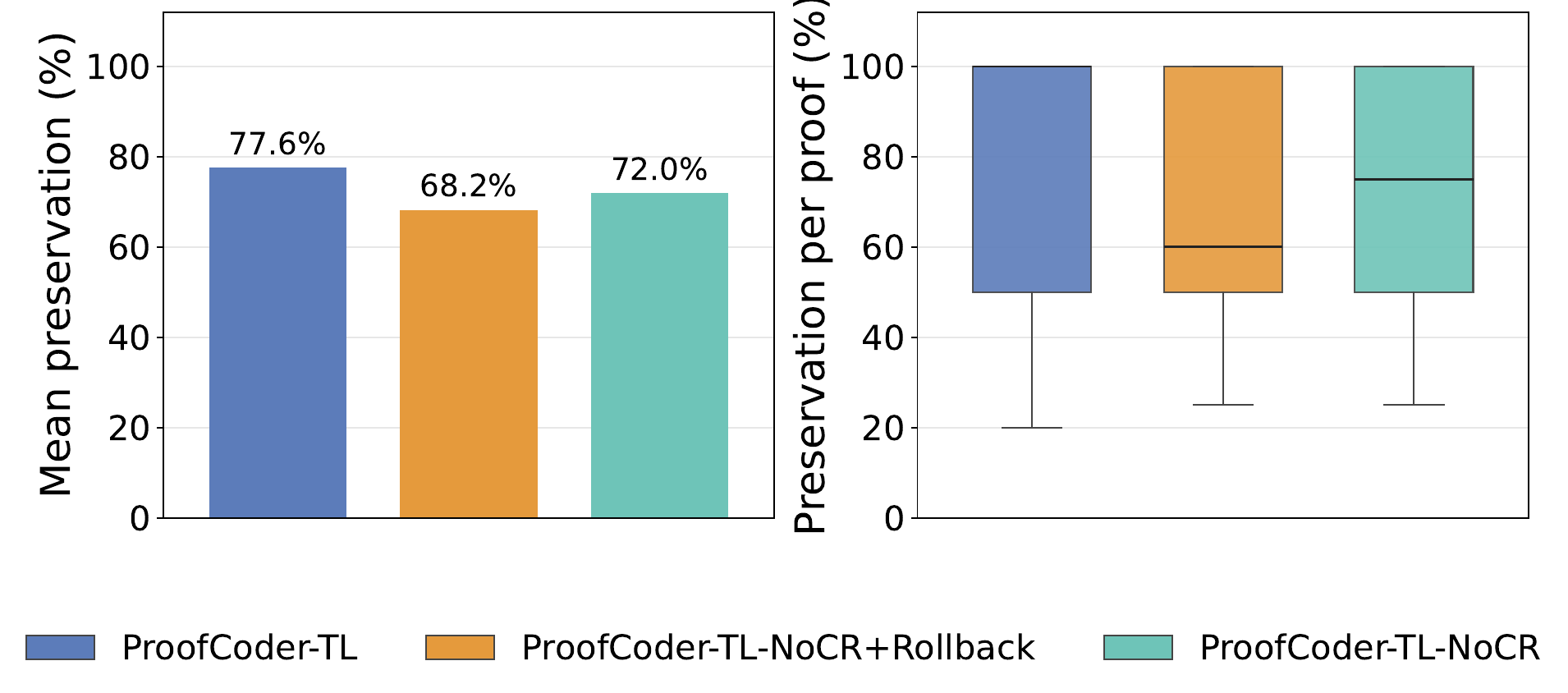}
}\hfill%
\caption{Backtrack preservation rate for the tactic-level ablations. Backtrack is considered preserved if the subsequent proof search is not entirely undone by the end of the run.}
\label{fig:backtrack_perservation}
\end{figure}

Figure~\ref{fig:backtrack_perservation} shows a consistent ordering on both benchmarks: \ProofCoderTL\ $>$ \ProofCoderTLNoCR\ $>$ \ProofCoderTLRollback. On SV-COMP, preservation drops from 88.7\% to 65.0\% without continuous reasoning and to 48.5\% with rollback, while the distributions widen substantially (interquartile ranges 80--100\%, 33--100\%, and 21--67\%, respectively). Verker shows the same ordering (77.6\%, 72.0\%, 68.2\%), but with smaller differences and largely overlapping distributions. This complements the depth and frequency results: \ProofCoderTLNoCR's shallow backtracks are more often overridden by later revisions, whereas \ProofCoderTLRollback's near-complete restarts discard earlier revisions by construction. Since the final backtrack is always preserved, \ProofCoderTL's higher rate also partly reflects its lower backtrack frequency. Overall, continuous reasoning reduces both the need for revision and the likelihood that revisions are later undone.


\RQAnswer{Answer to RQ2:}
{Continuous reasoning substantially changes backtracking behavior, mainly by reducing the need for revision: \ProofCoderTL\ backtracks approximately 4--7$\times$ less often than its ablations, and the revisions it does make are rarely undone. Without continuous reasoning, the agent performs frequent shallow backtracks that are later overridden by deeper ones, suggesting that continuous reasoning helps identify which proof prefix should be retained and which should be discarded. The interaction model also influences recovery strategy: adding the rollback tool leads to the most frequent and least stable backtracks, often near-complete restarts, indicating that multiple revision mechanisms lead the agent to discard its own repairs rather than refine them. Finally, across configurations, backtracking most frequently revises proof setup, particularly \texttt{intros}, followed by tactics such as \texttt{apply}, \texttt{assert}, and \texttt{destruct} that make substantive changes to the proof state. This suggests that agents use backtracking to revise higher-level proof decisions rather than simply regenerate failed tactics.}


\subsection{RQ3. How do different proof generation granularities compare?}

Prior work explores proof synthesis at different granularities, from tactic-level generation~\cite{rango, copra, prooverbot9001, pact} to whole-proof generation~\cite{baldur, deepseekprover1.5, goedelprover}. Tactic-level generation provides intermediate proof states for early error detection and adaptive search, while whole-proof generation reduces proof-assistant interactions. Direct comparison can be confounded by training-data contamination, as whole-proof generation may benefit from reproducing known proofs. AutoRocq benchmark mitigates this concern because its proof obligations have no ground-truth proofs available online. We therefore compare \ProofCoderTL\ and \ProofCoderWP\ within the same agent framework to isolate the effect of proof-generation granularity.


As discussed in RQ1 and shown in Tables~\ref{tab:svcomp-comparison} and \ref{tab:verker-comparison}, whole-proof generation tends to achieve slightly better overall performance. We next compare both configurations in terms of tool calls and backtracking, proof length, and proof similarity.
\vskip 0.3em
\noindent\noindent
\emph{{\textbf{Tool calls and backtracking.}}}
The tool-call traces show that \ProofCoderWP\ relies more heavily on explicit state inspection, particularly with \sol. On SV-COMP, \ProofCoderWP\ and \ProofCoderTL\ make 567 and 521 \texttt{check\_state} calls with \sonnet, but 206 and only 6 with \sol. The same pattern holds on Verker, where the configurations make 203 vs.\ 107 calls with \sonnet\ and 98 vs.\ 14 with \sol. This reflects their different feedback granularity: \ProofCoderWP\ receives the proof state only after each generated fragment and therefore uses \texttt{check\_state} to inspect intermediate states, whereas \ProofCoderTL\ receives the state after every tactic. The other context-retrieval actions remain broadly similar within each model: \texttt{read\_proof\_file} is comparable across configurations (131--148 with \sonnet; 184--189 with \sol), while \texttt{call\_command} is frequent for both and, on Verker, more frequent for \ProofCoderTL\ with \sol\ (695 vs.\ 483).

Backtracking shows the same configuration- and model-dependent pattern. On SV-COMP, \ProofCoderWP\ backtracks on 70.0\% of \sonnet\ lemmas (140/200), with 5.35 backtracks on average, compared with 65.0\% (130/200) and 3.69 for \ProofCoderTL; with \sol, these rates fall to 58.5\% and 26.5\%, with means of 1.01 and 0.45, respectively. Verker exhibits more frequent backtracking overall: with \sonnet, the corresponding rates are 91.2\% (73/80) and 86.2\% (69/80), with means of 10.89 and 6.04; with \sol, they are 68.8\% and 51.2\%, with means of 2.16 and 1.39. Thus, \ProofCoderWP\ consistently performs more explicit state inspections and backtracks than \ProofCoderTL, with both differences substantially larger for \sol.

\begin{figure}[htbp]
\resizebox{0.7\linewidth}{!}{%
\begin{minipage}{\linewidth}
\centering
\subfloat[Proof length (SV-COMP, Sonnet 4.5).\label{fig:svcomp_sonnet_length_hist}]{%
  \includegraphics[width=0.46\linewidth]{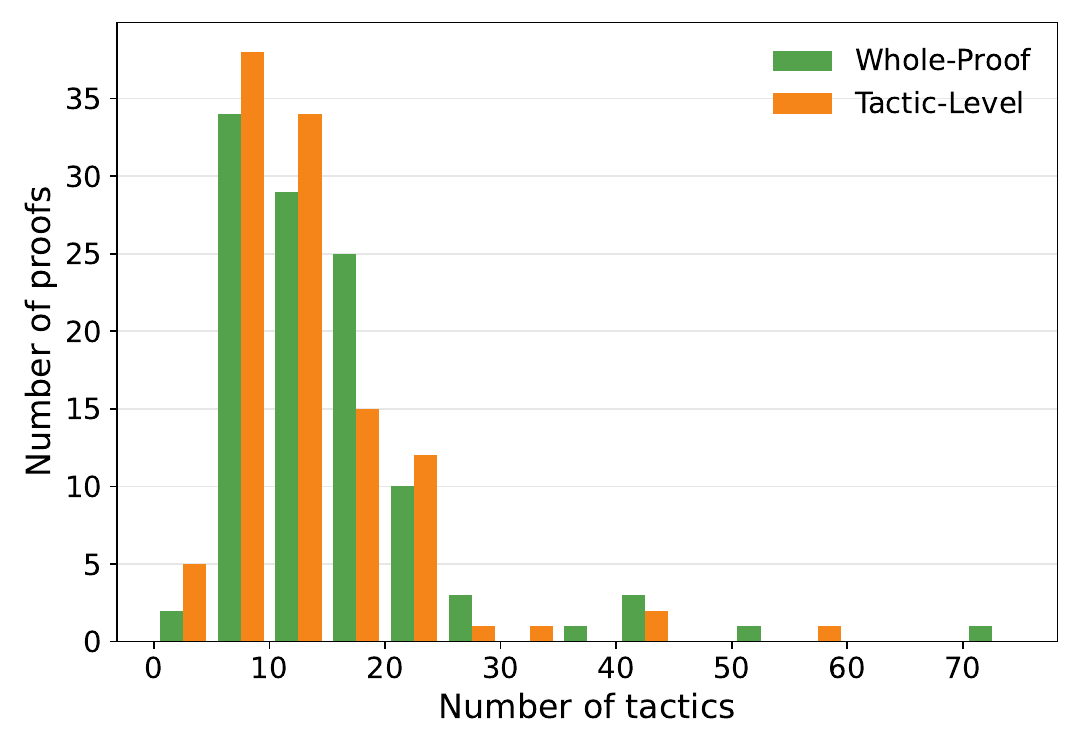}
}\hfill%
\subfloat[Proof similarity (SV-COMP, Sonnet 4.5). \label{fig:svcomp_sonnet_jaccard_hist}]{%
  \includegraphics[width=0.46\linewidth]{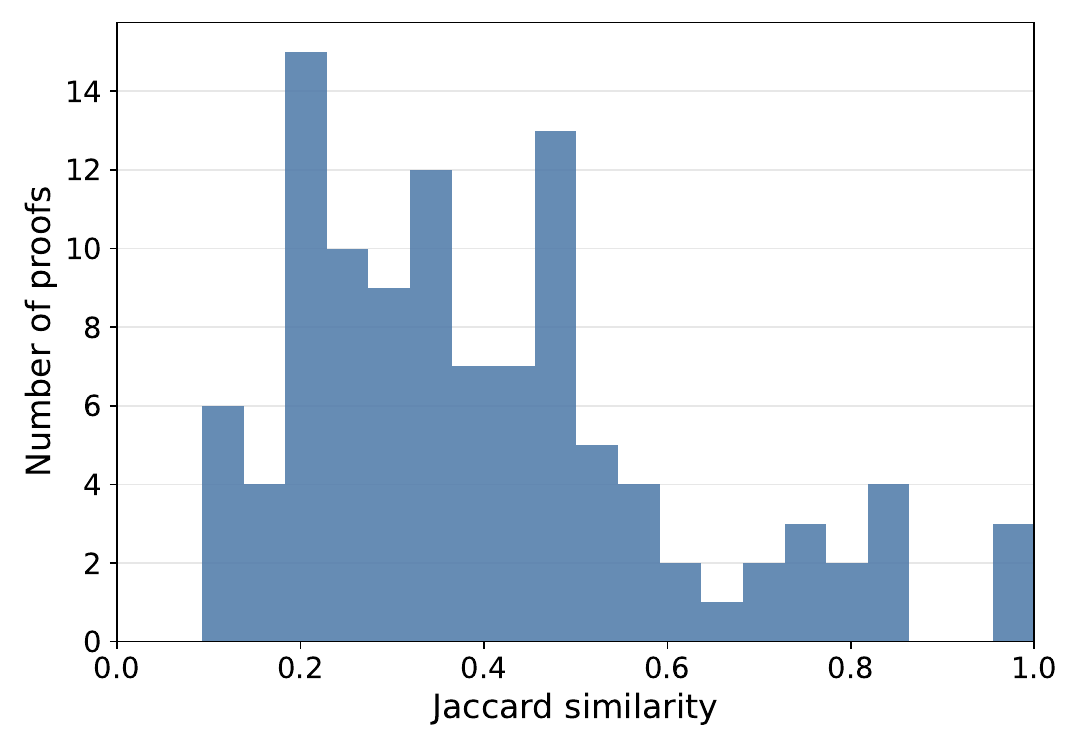}%
}\hfill
\subfloat[Proof length (SV-COMP, \sol). \label{fig:svcomp_sol_length_hist}]{%
  \includegraphics[width=0.46\linewidth]{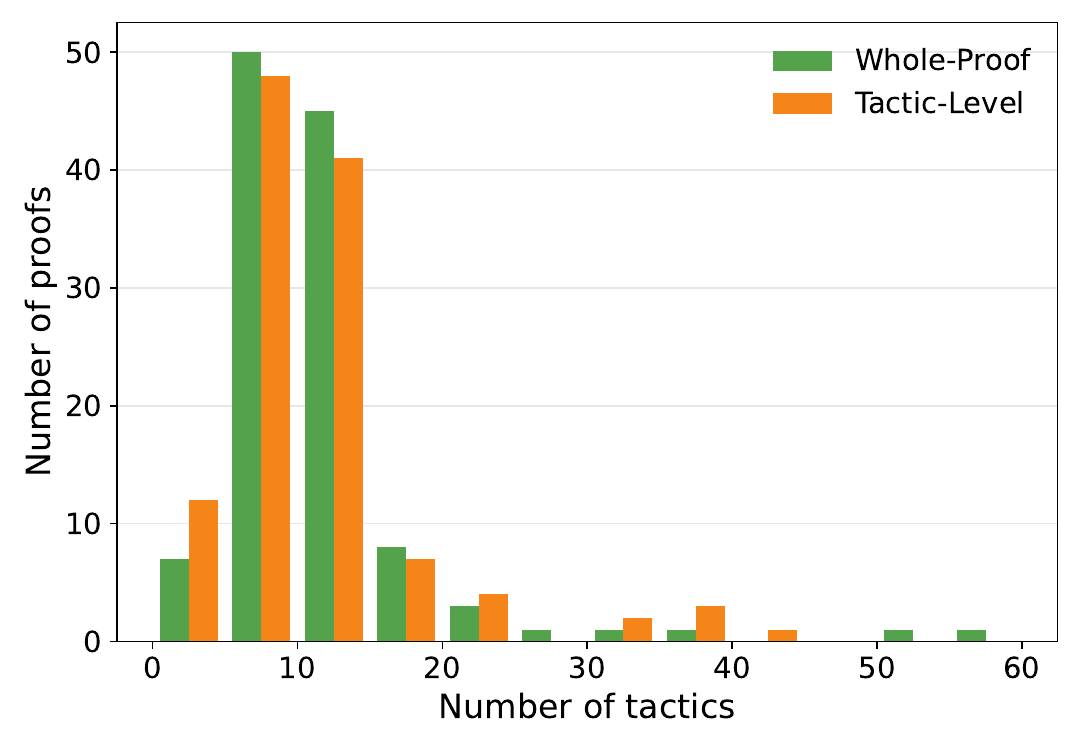}%
}\hfill
\subfloat[Proof similarity (SV-COMP, \sol). \label{fig:svcomp_sol_jaccard_hist}]{%
  \includegraphics[width=0.46\linewidth]{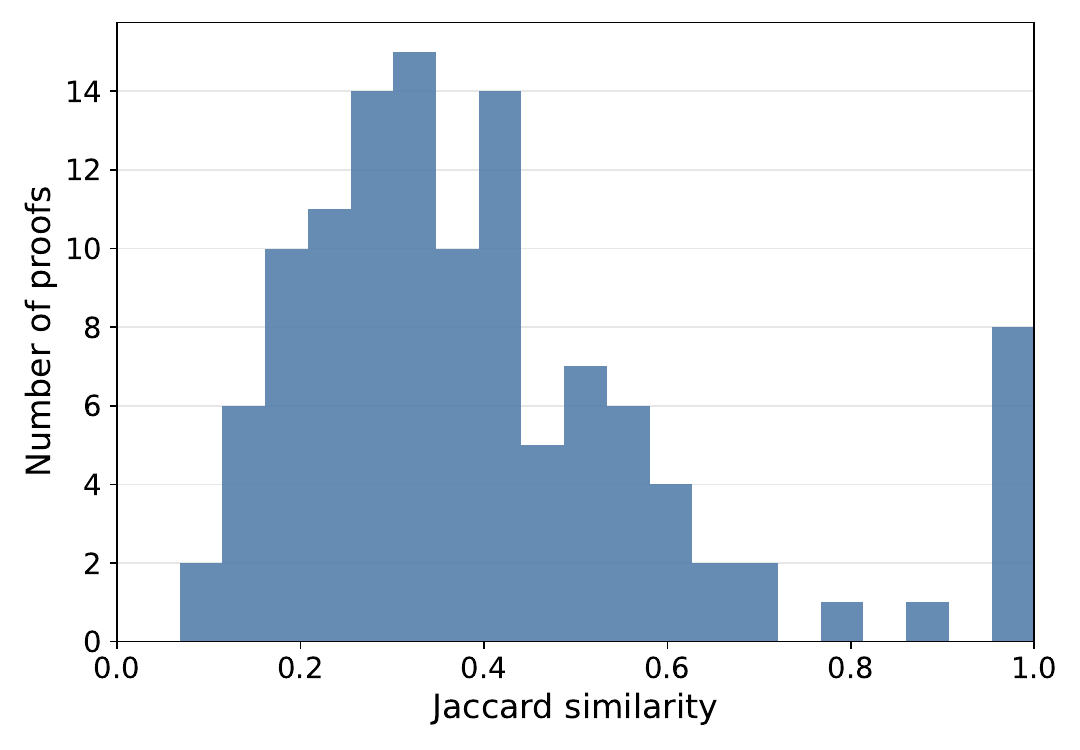}%
}\hfill

\subfloat[Proof length (Verker, Sonnet 4.5). \label{fig:verker_sonnet_length_hist}]{%
  \includegraphics[width=0.46\linewidth]{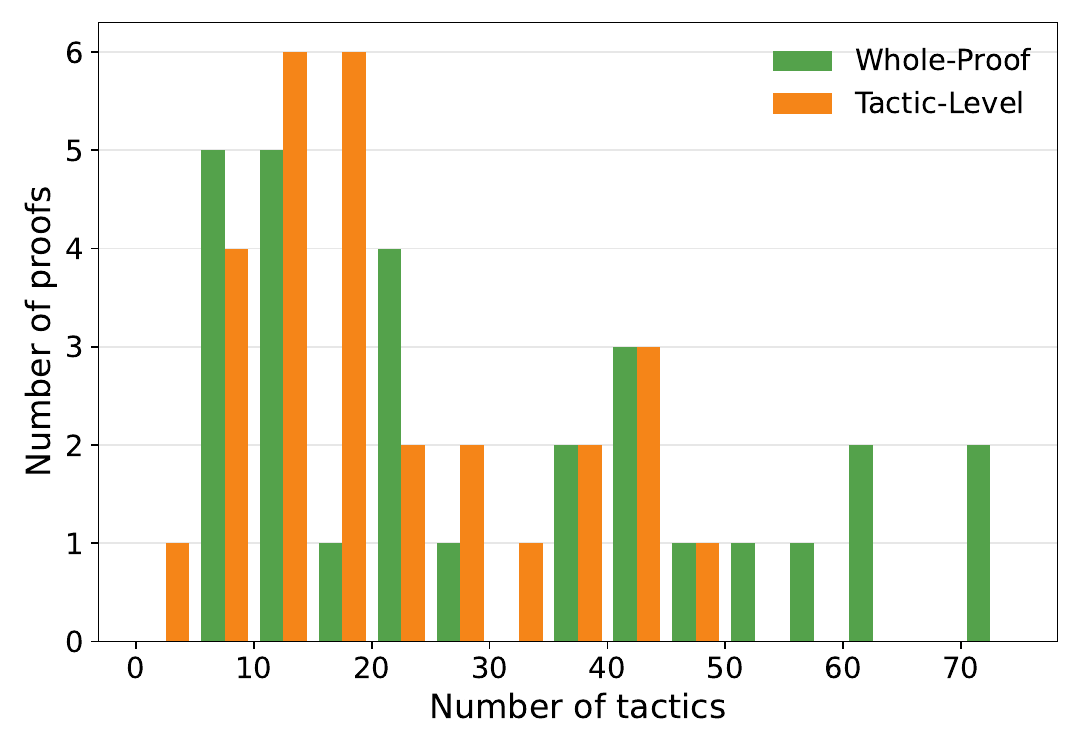}%
}\hfill
\subfloat[Proof similarity (Verker, Sonnet 4.5). \label{fig:verker_sonnet_jaccard_hist}]{%
  \includegraphics[width=0.46\linewidth]{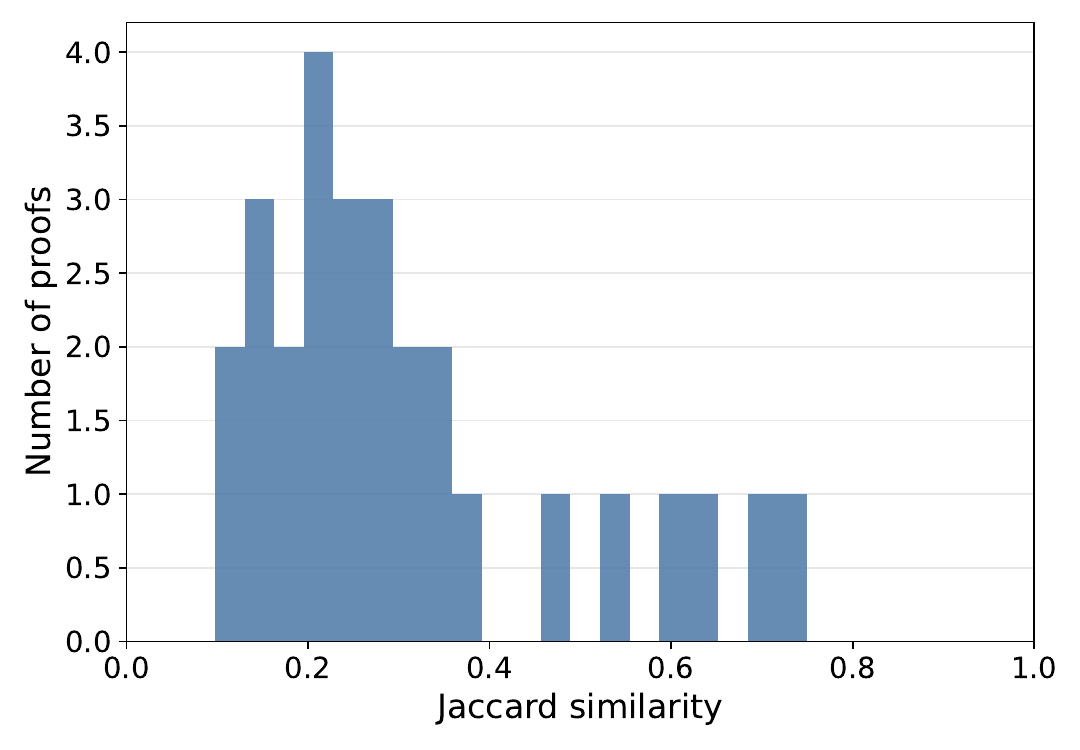}%
}\hfill
\subfloat[Proof length (Verker, \sol). \label{fig:verker_sol_length_hist}]{%
  \includegraphics[width=0.46\linewidth]{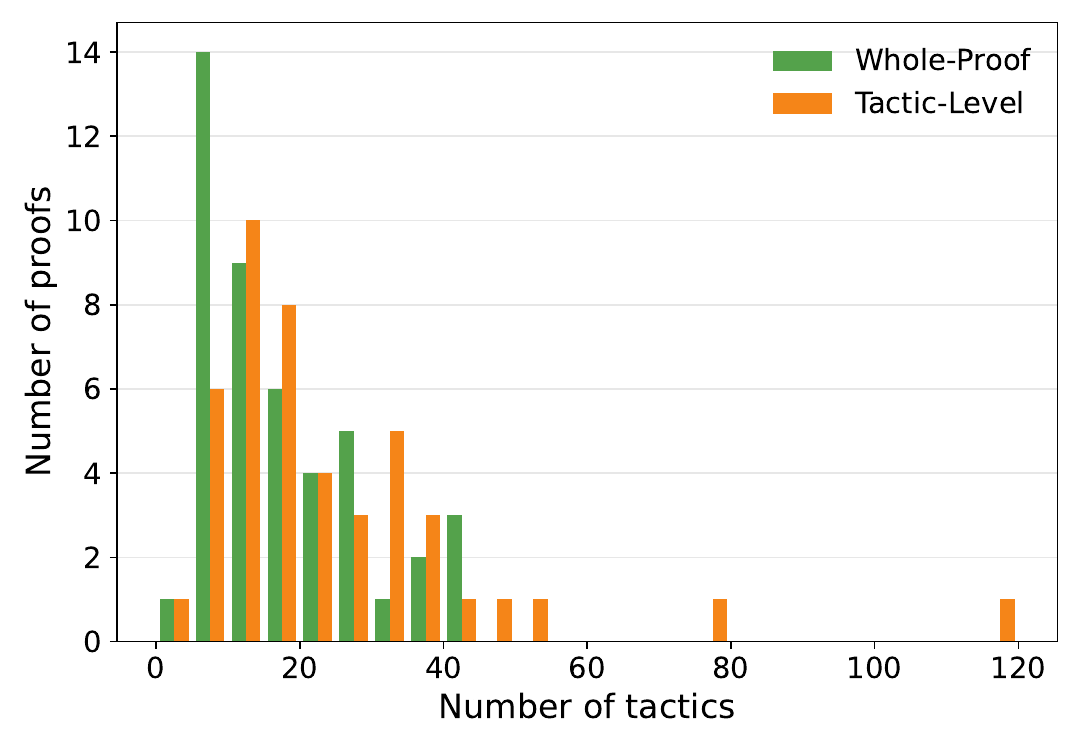}%
}\hfill
\subfloat[Proof similarity (Verker, \sol). \label{fig:verker_sol_jaccard_hist}]{%
  \includegraphics[width=0.46\linewidth]{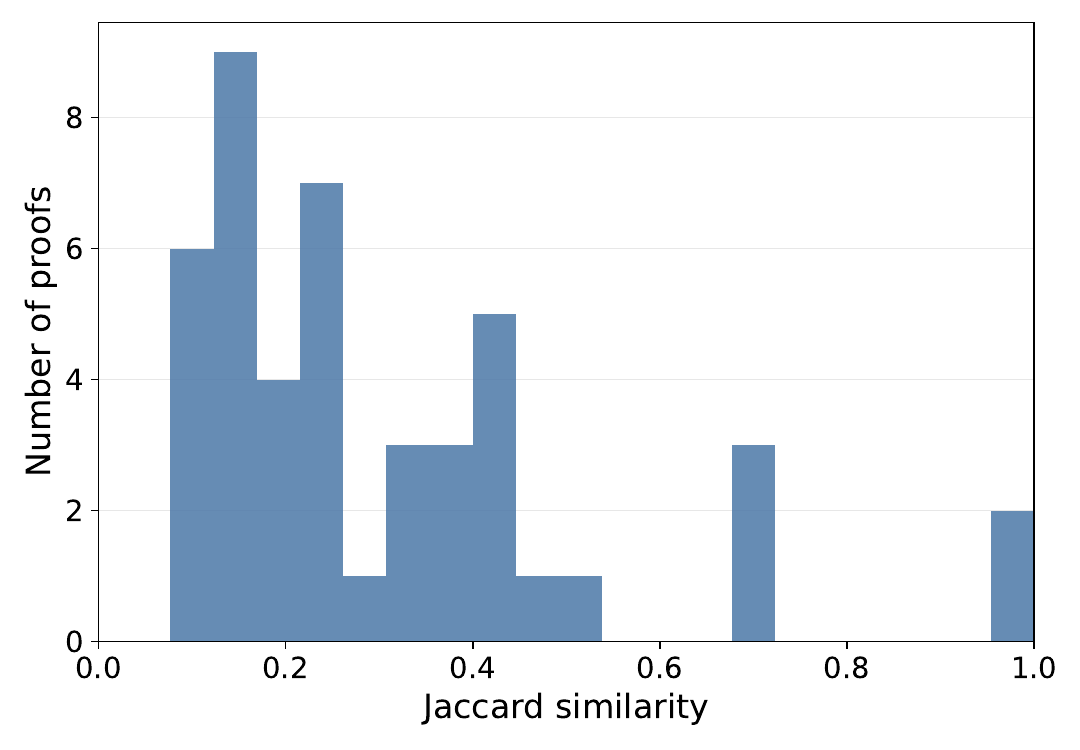}%
}\hfill
\end{minipage}
}
\caption{Comparison of length and similarity between proofs generated by \ProofCoderTL\ and \ProofCoderWP.}
\label{fig:proof_comparison}
\end{figure}

\vskip 0.3em
\noindent\noindent
\emph{{\textbf{Proof length and similarity.}}} 
Figure~\ref{fig:proof_comparison} compares proofs produced by \ProofCoderTL\ and \ProofCoderWP\ on lemmas solved by both configurations. This intersection contains 109 and 118 SV-COMP lemmas for \sonnet\ and \sol, respectively, and 28 and 45 Verker lemmas. Before computing Jaccard similarity, we remove comments and canonicalize introduced and generated variables, and numeric literals.

On SV-COMP, proofs are generally short and similar in length across configurations. With \sonnet, \ProofCoderWP\ and \ProofCoderTL\ average 14.7 and 12.9 tactics, respectively; with \sol, they average 10.8 and 10.6. At least 93\% contain 25 tactics or fewer, although \ProofCoderWP\ exhibits a somewhat longer tail. Proof similarity is moderate, with mean Jaccard scores of 0.41 for \sonnet\ and 0.40 for \sol. Only 3 and 8 proof pairs, respectively, are identical, indicating that the two configurations generally construct distinct proofs despite sharing many tactic steps.

On Verker, proofs are longer and the effect of granularity depends on the model. When using \sonnet, \ProofCoderWP\ averages 30.4 tactics compared with 20.8 for \ProofCoderTL; with \sol, the pattern reverses, with averages of 16.7 and 23.8. Similarity is also lower than on SV-COMP, with mean Jaccard scores of 0.30 and 0.31 for \sonnet\ and \sol, respectively. Across both models, only two proofs are identical with \sol.

Taken together, the results show no consistent advantage of either granularity in proof length, while the relatively low similarity indicates that granularity materially affects how proofs are constructed. Combined with the tool-call traces, this suggests different search and repair strategies: \ProofCoderWP\ relies more on explicit state inspection and backtracking, whereas \ProofCoderTL\ receives feedback after every tactic and therefore requires fewer explicit revisions.

\RQAnswer{Answer to RQ3:}
{Whole-proof and tactic-level generation achieve similar proof success, differing by at most three lemmas per setting, with \ProofCoderWP\ ahead in three of four. Granularity mainly affects interaction behavior: as \ProofCoderWP\ receives feedback only after each generated fragment, it performs more explicit state inspections (by up to 34$\times$ with \sol) and backtracks more often, while \ProofCoderTL\ receives the proof state after every tactic, requiring fewer explicit checks and revisions. Despite this, \ProofCoderWP\ is cheaper in all settings, as larger fragments require fewer model turns. On lemmas solved by both configurations, proofs have moderate tactic overlap (mean Jaccard 0.30--0.41), are rarely identical, and neither granularity consistently yields shorter proofs. Granularity has little effect on success rate but substantially shapes search, repair, and the resulting proof scripts.}

%
\section{Related Work}
Deep learning has become prominent in theorem proving~\cite{surveyNTP, rango, baldur, copra, legoprover, dsp, leandojo, qedcartographer, cobblestone, proofcoop, prooverbot9001, tactok, coqgym, poetry, magnushammer, lyra, subgoal, seed-prover, fvel, autorocq, autorocq_lemma, graph2tac, diva} and software verification~\cite{ai-assisted_dafny, code2inv, LearningLoopInv, autorocq, llm_model_checking, pat-agent, concolicexec, autorocq_lemma, specgen, vecogen, fvel, specrover}. Recent work on automated theorem proving for software verification has particularly benefited from LLMs. 

Rango~\cite{rango} fine-tunes an LLM for tactic-level Rocq proof generation: at each step, the model receives proof-assistant context, relevant lemmas, and premises, and generates the next tactic. When a tactic fails, Rango recovers by restarting proof search from the beginning. COPRA~\cite{copra} also performs tactic-level generation, prompting the model with the current proof state and additional context at each step. However, backtracking is controlled by an external depth-first search procedure: once the allowed attempts from a state are exhausted, the controller returns to an earlier proof state and asks the model for a different tactic.
PALM~\cite{palm} follows a generate-then-repair pipeline: it retrieves relevant lemmas and definitions, prompts the LLM to generate a proof script, executes it, and applies symbolic repairs when errors occur. If repair fails, PALM backtracks and attempts to solve earlier goals using CoqHammer~\cite{coqhammer}. 
Cobblestone~\cite{cobblestone} uses a divide-and-conquer strategy. It executes LLM-generated proof scripts in a fail-safe mode, preserves the fragments accepted by Rocq, replaces failing parts with temporary admits, and recursively proves the remaining subgoals. The final assembled proof is accepted only after Rocq checking, often with support from tools such as CoqHammer.
Baldur~\cite{baldur} explores whole-proof generation for Isabelle/HOL, using LLMs to produce complete proof scripts rather than individual tactics. When a proof fails, Baldur conditions a learned repair model on the failed script and proof-assistant error message to revise local mistakes.
In these systems, the model is primarily invoked at a specific step to generate tactics, proof scripts, or proof fragments, while search control, retrieval, and recovery are largely managed by the surrounding algorithm in a deterministic, yet more restrictive manner.

%
AutoRocq~\cite{autorocq} is, to our knowledge, the state-of-the-art agentic approach for Rocq-based software verification and an important step toward more autonomous proof synthesis. It equips the LLM with tools for context retrieval, planning, tactic generation, and rollback, and shows that this agentic formulation can substantially outperform earlier approaches. AutoRocq nevertheless retains a structured interaction policy, with prompting that guides tool selection, limits repeated queries, and encourages particular actions in specific situations.
\toolname\ proposes a less prescribed interaction model, allowing the agent to freely call and combine tools and to revise proofs through a compact editing and backtracking mechanism.

\section{Threats to Validity}
\noindent{\emph{\textbf{Construct validity.}}}
Comparing whole-proof and tactic-level approaches introduces inherent differences in resource consumption, as whole-proof generation typically requires fewer LLM calls. Using the number of LLM calls as the stopping criterion would therefore systematically favor whole-proof generation. To mitigate this threat, we use the total number of non-cached LLM tokens as the primary stopping criterion, providing a more comparable computational budget across approaches. We additionally impose maximum wall-clock time and total token-budget limits. Preliminary experiments showed that increasing these limits yielded only marginal improvements in proof success, suggesting that performance had largely saturated under the allocated resources.

\noindent{\emph{\textbf{Internal validity.}}}
Although both \toolname\ and AutoRocq are agentic theorem provers that interact with Rocq through CoqPyt, they use different agent frameworks. AutoRocq relies on LiteLLM~\cite{litellm}, whereas \toolname\ is a thin integration layer over OpenCode. Consequently, framework differences in context management, conversation history, or tool-call serialization may influence performance. To mitigate this threat, we complement the comparison with an ablation study in which all \toolname\ variants use the same OpenCode-based infrastructure, isolating the impact of the proposed design choices.

\noindent{\emph{\textbf{External validity.}}}
Due to resource constraints, we do not evaluate on the full AutoRocq benchmark. Since there are no ground-truth proofs, evaluating every obligation would require spending a substantial fraction of the token budget on obligations that may not be provable. Instead, we evaluate the 200 SV-COMP obligations with the highest term complexity, yielding a challenging yet tractable subset. To ensure that this subset contains a substantial number of provable obligations, we confirmed that 88 were solved in the original AutoRocq evaluation by either AutoRocq or one of its baselines. Thus, our findings should be interpreted as evidence of \toolname's effectiveness on this challenging subset rather than as a complete characterization of AutoRocq’s benchmark.


\section{Conclusion}




We presented \toolname, an agentic Rocq prover that advances the state of the art on challenging verification benchmarks while introducing a new perspective on LLM-based theorem proving. Our results demonstrate that the design of the agent interface is an important factor in proving performance. In particular, continuous reasoning and structured proof revision substantially improve an agent's ability to recover from failed proof attempts. These findings suggest that LLM-based theorem provers should be designed not only around tactic generation, but also around feedback-driven proof repair. More broadly, our work indicates that continuous reasoning and robust backtracking policies are key ingredients for building effective agentic verification systems.

Future work includes
evaluating these ideas on additional proof assistants and verification-aware
languages, as well as exploring richer verifier interactions and collaborative
multi-agent proof search.

\section{Data Availability}
We make all code, experimental scripts, and data publicly available: 
\url{https://doi.org/10.5281/zenodo.21084811}

\section{Acknowledgments}
This work was supported by national funds through Fundação para a Ciência e a Tecnologia, I.P. (FCT) under projects UID/50021/2025 (DOI: \href{https://doi.org/10.54499/UID/50021/2025}{10.54499/UID/50021/2025}), UID/PRR/50021/2025 (DOI: \href{https://doi.org/10.54499/UID/PRR/50021/2025}{UID/ PRR/50021/2025}), and
 co-funded by the European Union through the Lisboa 2030 Programme (ERDF) and by national funds through FCT, I.P., under project no. 15018 (project SafeIaC, DOI: \href{https://doi.org/10.54499/2023.18089.ICDT}{10.54499/ 2023.18089.ICDT}).
 We thank Amazon Science for the Amazon Research Awards (Fall 2024), which enabled the experiments through access to AWS Bedrock.

\bibliographystyle{ACM-Reference-Format}
\bibliography{sample-base}

@String{Computing = "Computing" }

@String{Computer = "{IEEE} Computer" }

@String{Springer = "Springer-Verlag" }

@article{autorocq,
  title={Agentic Verification of Software Systems},
  author={Tu, Haoxin and Zhao, Huan and Song, Yahui and Zafar, Mehtab and Meng, Ruijie and Roychoudhury, Abhik},
  year = {2026},
  issue_date = {July 2026},
  publisher = {Association for Computing Machinery},
  address = {New York, NY, USA},
  volume = {3},
  number = {FSE},
  url = {https://doi.org/10.1145/3808164},
  doi = {10.1145/3808164},
  journal = {Proc. ACM Softw. Eng.}
}

@article{autorocq_lemma,
  title={Lemma Discovery in Agentic Program Verification},
  author={Zhao, Huan and Tu, Haoxin and Liu, Zhengyao and Rinard, Martin and Roychoudhury, Abhik},
  journal={arXiv preprint arXiv:2603.22114},
  year={2026}
}

@inproceedings{cobblestone,
  author = {Saketh Ram Kasibatla and Arpan Agrawal and Yuriy Brun and 
            Sorin Lerner and Talia Ringer and Emily First},
  title =
  {Cobblestone: 
  A Divide-and-Conquer Approach for Automating Formal Verification},
  booktitle = {Proceedings of the 48th International Conference on Software Engineering (ICSE)},
  venue = {ICSE},
  address = {Rio de Janeiro, Brazil},
  month = {April},
  year = {2026},
  pages = {714--726},
  doi = {10.1145/3744916.3773178},
}

@inproceedings{palm,
  title={Proof automation with large language models},
  author={Lu, Minghai and Delaware, Benjamin and Zhang, Tianyi},
  booktitle={Proceedings of the 39th IEEE/ACM International Conference on Automated Software Engineering},
  pages={1509--1520},
  year={2024}
}

@inproceedings{rango,
  title={Rango: Adaptive Retrieval-Augmented Proving for Automated Software Verification},
  author={Thompson, Kyle and Saavedra, Nuno and Carrott, Pedro and Fisher, Kevin and Sanchez-Stern, Alex and Brun, Yuriy and Ferreira, Jo{\~a}o F and Lerner, Sorin and First, Emily},
  booktitle={2025 IEEE/ACM 47th International Conference on Software Engineering (ICSE)},
  pages={347--359},
  year={2025},
  organization={IEEE}
}

@inproceedings{
copra,
title={An In-Context Learning Agent for Formal Theorem-Proving},
author={Amitayush Thakur and George Tsoukalas and Yeming Wen and Jimmy Xin and Swarat Chaudhuri},
booktitle={First Conference on Language Modeling},
year={2024},
}

@inproceedings{filliatre2013why3,
  title={Why3---where programs meet provers},
  author={Filli{\^a}tre, Jean-Christophe and Paskevich, Andrei},
  booktitle={European symposium on programming},
  pages={125--128},
  year={2013},
  organization={Springer}
}

@article{kirchner2015frama,
  title={Frama-C: A software analysis perspective},
  author={Kirchner, Florent and Kosmatov, Nikolai and Prevosto, Virgile and Signoles, Julien and Yakobowski, Boris},
  journal={Formal aspects of computing},
  volume={27},
  number={3},
  pages={573--609},
  year={2015},
  publisher={Springer}
}

@incollection{acsl,
  title={Formally expressing what a program should do: the ACSL language},
  author={Blanchard, Allan and March{\'e}, Claude and Prevosto, Virgile},
  booktitle={Guide to Software Verification with Frama-C: Core Components, Usages, and Applications},
  pages={3--80},
  year={2024},
  publisher={Springer}
}

@inproceedings{
surveyNTP,
title={A Survey on Deep Learning for Theorem Proving},
author={Zhaoyu Li and Jialiang Sun and Logan Murphy and Qidong Su and Zenan Li and Xian Zhang and Kaiyu Yang and Xujie Si},
booktitle={First Conference on Language Modeling},
year={2024},
}

@article{leandojo,
  title={LeanDojo: Theorem proving with retrieval-augmented language models},
  author={Yang, Kaiyu and Swope, Aidan and Gu, Alex and Chalamala, Rahul and Song, Peiyang and Yu, Shixing and Godil, Saad and Prenger, Ryan J and Anandkumar, Animashree},
  journal={Advances in Neural Information Processing Systems},
  volume={36},
  pages={21573--21612},
  year={2023}
}

@inproceedings{
dsp,
title={{Draft}, {Sketch}, and {Prove}: {Guiding} Formal Theorem Provers with Informal Proofs},
author={Albert Qiaochu Jiang and Sean Welleck and Jin Peng Zhou and Wenda Li and Jiacheng Liu and Mateja Jamnik and Timoth{\'{e}}e Lacroix and Yuhuai Wu and Guillaume Lample},
booktitle={International Conference on Learning Representations},
year={2023},
url={https://doi.org/10.48550/arXiv.2210.12283}
}

@inproceedings{baldur,
  title={Baldur: Whole-proof generation and repair with large language models},
  author={First, Emily and Rabe, Markus N and Ringer, Talia and Brun, Yuriy},
  booktitle={Proceedings of the 31st ACM Joint European Software Engineering Conference and Symposium on the Foundations of Software Engineering},
  pages={1229--1241},
  year={2023}
}

@inproceedings{legoprover,
  title={Lego-prover: Neural theorem proving with growing libraries},
  author={Wang, Haiming and Xin, Huajian and Zheng, Chuanyang and Liu, Zhengying and Cao, Qingxing and Huang, Yinya and Xiong, Jing and Shi, Han and Xie, Enze and Yin, Jian and others},
  booktitle={International Conference on Learning Representations},
  volume={2024},
  pages={31566--31597},
  year={2024}
}

@article{tactok,
  title={TacTok: Semantics-aware proof synthesis},
  author={First, Emily and Brun, Yuriy and Guha, Arjun},
  journal={Proceedings of the ACM on Programming Languages},
  volume={4},
  number={OOPSLA},
  pages={1--31},
  year={2020},
  publisher={ACM New York, NY, USA}
}

@inproceedings{coqgym,
  title={Learning to prove theorems via interacting with proof assistants},
  author={Yang, Kaiyu and Deng, Jia},
  booktitle={International Conference on Machine Learning},
  pages={6984--6994},
  year={2019},
  organization={PMLR}
}

@inproceedings{prooverbot9001,
  title={Generating correctness proofs with neural networks},
  author={Sanchez-Stern, Alex and Alhessi, Yousef and Saul, Lawrence and Lerner, Sorin},
  booktitle={Proceedings of the 4th ACM SIGPLAN International Workshop on Machine Learning and Programming Languages},
  pages={1--10},
  year={2020}
}

@article{software_quality,
  title={Automatically Engineering Trusted Software: A Research Roadmap},
  author={Brun, Yuriy and Chakraborty, Saikat and Le Goues, Claire and P{\u{a}}s{\u{a}}reanu, Corina and Singla, Adish},
  journal={ACM Transactions on Software Engineering and Methodology},
  year={2026},
  publisher={ACM New York, NY}
}

@inproceedings{diva,
  title={Diversity-driven automated formal verification},
  author={First, Emily and Brun, Yuriy},
  booktitle={Proceedings of the 44th International Conference on Software Engineering},
  pages={749--761},
  year={2022}
}

@inproceedings{specrover,
  title={Specrover: Code intent extraction via llms},
  author={Ruan, Haifeng and Zhang, Yuntong and Roychoudhury, Abhik},
  booktitle={2025 IEEE/ACM 47th International Conference on Software Engineering (ICSE)},
  pages={963--974},
  year={2025},
  organization={IEEE}
}

@article{graph2tac,
  title={Graph2Tac: Online representation learning of formal math concepts},
  author={Blaauwbroek, Lasse and Ol{\v{s}}{\'a}k, Miroslav and Rute, Jason and Massolo, Fidel Ivan Schaposnik and Piepenbrock, Jelle and Pestun, Vasily},
  journal={arXiv preprint arXiv:2401.02949},
  year={2024}
}

@article{fv_survey,
  title={QED at large: A survey of engineering of formally verified software},
  author={Ringer, Talia and Palmskog, Karl and Sergey, Ilya and Milos, Gligoric and Tatlock, Zachary},
  journal={Foundations and Trends in Programming Languages},
  volume={5},
  number={2-3},
  pages={102--281},
  year={2019},
  publisher={Emerald Publishing Limited}
}

@article{fvel,
  title={FVEL: Interactive Formal Verification Environment with Large Language Models via Theorem Proving},
  author={Lin, Xiaohan and Cao, Qingxing and Huang, Yinya and Wang, Haiming and Lu, Jianqiao and Liu, Zhengying and Song, Linqi and Liang, Xiaodan},
  journal={Advances in Neural Information Processing Systems},
  volume={37},
  pages={54932--54946},
  year={2024}
}

@inproceedings{vecogen,
  title={VecoGen: Automating Generation of Formally Verified C Code with Large Language Models},
  author={Sevenhuijsen, Merlijn and Etemadi, Khashayar and Nyberg, Mattias},
  booktitle={2025 IEEE/ACM 13th International Conference on Formal Methods in Software Engineering (FormaliSE)},
  pages={101--112},
  year={2025},
  organization={IEEE}
}

@inproceedings{specgen,
  title={SpecGen: Automated Generation of Formal Program Specifications via large language models},
  author={Ma, Lezhi and Liu, Shangqing and Li, Yi and Xie, Xiaofei and Bu, Lei},
  booktitle={2025 IEEE/ACM 47th International Conference on Software Engineering (ICSE)},
  pages={16--28},
  year={2025},
  organization={IEEE}
}

@article{seed-prover,
  title={{Seed-Prover}: Deep and broad reasoning for automated theorem proving},
  author={Chen, Luoxin and Gu, Jinming and Huang, Liankai and Huang, Wenhao and Jiang, Zhicheng and Jie, Allan and Jin, Xiaoran and Jin, Xing and Li, Chenggang and Ma, Kaijing and others},
  journal={arXiv preprint arXiv:2507.23726},
  year={2025}
}

@article{lyra,
  title={Lyra: Orchestrating dual correction in automated theorem proving},
  author={Zheng, Chuanyang and Wang, Haiming and Xie, Enze and Liu, Zhengying and Sun, Jiankai and Xin, Huajian and Shen, Jianhao and Li, Zhenguo and Li, Yu},
  journal={arXiv preprint arXiv:2309.15806},
  year={2023}
}

@inproceedings{subgoal,
  title={Subgoal-based demonstration learning for formal theorem proving},
  author={Zhao, Xueliang and Li, Wenda and Kong, Lingpeng},
  booktitle={Forty-first International Conference on Machine Learning},
  year={2024}
}

@inproceedings{magnushammer,
  title={Magnushammer: A transformer-based approach to premise selection},
  author={Miku{\l}a, Maciej and Tworkowski, Szymon and Antoniak, Szymon and Piotrowski, Bartosz and Jiang, Qiaochu and Zhou, Jin and Szegedy, Christian and Kuci{\'n}ski, {\L}ukasz and Mi{\l}o{\'s}, Piotr and Wu, Yuhuai},
  booktitle={International Conference on Learning Representations},
  volume={2024},
  pages={39326--39350},
  year={2024}
}

@article{poetry,
  title={Proving theorems recursively},
  author={Wang, Haiming and Xin, Huajian and Liu, Zhengying and Li, Wenda and Huang, Yinya and Lu, Jianqiao and Yang, Zhicheng and Tang, Jing and Yin, Jian and Li, Zhenguo and others},
  journal={Advances in Neural Information Processing Systems},
  volume={37},
  pages={86720--86748},
  year={2024}
}

@inproceedings{qedcartographer,
  title={QEDCartographer: Automating formal verification using reward-free reinforcement learning},
  author={Sanchez-Stern, Alex and Varghese, Abhishek and Kaufman, Zhanna and Zhang, Dylan and Ringer, Talia and Brun, Yuriy},
  booktitle={2025 IEEE/ACM 47th International Conference on Software Engineering (ICSE)},
  pages={307--320},
  year={2025},
  organization={IEEE}
}

@inproceedings{CoqPyt,
  title={CoqPyt: Proof navigation in Python in the era of LLMs},
  author={Carrott, Pedro and Saavedra, Nuno and Thompson, Kyle and Lerner, Sorin and Ferreira, Jo{\~a}o F and First, Emily},
  booktitle={Companion Proceedings of the 32nd ACM International Conference on the Foundations of Software Engineering},
  pages={637--641},
  year={2024}
}

@inproceedings{proofcoop,
  title={ProofCoop: Collaborative Automated Formal Verification},
  author={Kaufman, Zhanna and First, Emily and Sanchez-Stern, Alex and Thompson, Kyle and Lerner, Sorin and Brun, Yuriy},
  booktitle={Proceedings of the 48th International Conference on Software Engineering (ICSE). Rio de Janeiro, Brazil},
  year={2026}
}

@article{LearningLoopInv,
  title={Learning loop invariants for program verification},
  author={Si, Xujie and Dai, Hanjun and Raghothaman, Mukund and Naik, Mayur and Song, Le},
  journal={Advances in Neural Information Processing Systems},
  volume={31},
  year={2018}
}

@inproceedings{code2inv,
  title={Code2inv: A deep learning framework for program verification},
  author={Si, Xujie and Naik, Aaditya and Dai, Hanjun and Naik, Mayur and Song, Le},
  booktitle={International Conference on Computer Aided Verification},
  pages={151--164},
  year={2020},
  organization={Springer}
}

@inproceedings{pat-agent,
  title={PAT-Agent: Autoformalization for Model Checking},
  author={Zuo, Xinyue and Zhang, Yifan and Wang, Hongshu and Cai, Yufan and Hou, Zhe and Sun, Jing and Dong, Jin Song},
  booktitle={2025 40th IEEE/ACM International Conference on Automated Software Engineering (ASE)},
  pages={2122--2133},
  year={2025},
  organization={IEEE}
}

@inproceedings{llm_model_checking,
  title={LLM Meets Bounded Model Checking: Neuro-Symbolic Loop Invariant Inference},
  author={Wu, Guangyuan and Cao, Weining and Yao, Yuan and Wei, Hengfeng and Chen, Taolue and Ma, Xiaoxing},
  booktitle={Proceedings of the 39th IEEE/ACM International Conference on Automated Software Engineering},
  pages={406--417},
  year={2024}
}

@article{ai-assisted_dafny,
  title={Towards ai-assisted synthesis of verified dafny methods},
  author={Misu, Md Rakib Hossain and Lopes, Cristina V and Ma, Iris and Noble, James},
  journal={Proceedings of the ACM on Software Engineering},
  volume={1},
  number={FSE},
  pages={812--835},
  year={2024},
  publisher={ACM New York, NY, USA}
}

@inproceedings{concolicexec,
  title={Agentic Concolic Execution},
  author={Luo, Zhengxiong and Zhao, Huan and Wolff, Dylan and Cadar, Cristian and Roychoudhury, Abhik},
  booktitle={Proceedings of the IEEE Symposium on Security and Privacy (S\&P)},
  pages={1--19},
  year={2026}
}

@article{sweagent,
  title={{SWE}-agent: Agent-computer interfaces enable automated software engineering},
  author={Yang, John and Jimenez, Carlos and Wettig, Alexander and Lieret, Kilian and Yao, Shunyu and Narasimhan, Karthik and Press, Ofir},
  journal={Advances in Neural Information Processing Systems},
  volume={37},
  pages={50528--50652},
  year={2024}
}

@inproceedings{react,
  author = {Yao, Shunyu and Zhao, Jeffrey and Yu, Dian and Du, Nan and Shafran, Izhak and Narasimhan, Karthik R. and Cao, Yuan},
  booktitle = {ICLR},
  ee = {https://openreview.net/forum?id=WE_vluYUL-X},
  publisher = {OpenReview.net},
  title = {ReAct: Synergizing Reasoning and Acting in Language Models},
  url = {http://dblp.uni-trier.de/db/conf/iclr/iclr2023.html#YaoZYDSN023},
  year = 2023
}

@article{coqhammer,
  title={Hammer for Coq: Automation for dependent type theory},
  author={Czajka, {\L}ukasz and Kaliszyk, Cezary},
  journal={Journal of automated reasoning},
  volume={61},
  number={1},
  pages={423--453},
  year={2018},
  publisher={Springer}
}

@misc{opencode,
  title={{OpenCode}},
  author={{Anomaly}},
  year={2025},
  howpublished={\url{https://github.com/anomalyco/opencode}}
}

@misc{litellm,
  title={{LiteLLM}: Achieving Sub-Millisecond Proxy Overhead},
  author={{BerriAI}},
  year={2025},
  howpublished={\url{https://docs.litellm.ai/blog/sub-millisecond-proxy-overhead}}
}

@misc{coq,
	title        = {Coq, v.8.7},
	author       = {{The Coq Development Team}},
	year         = 2017,
	howpublished = {\url{https://coq.inria.fr}},
}

@inproceedings{lean,
  author    = {de Moura, Leonardo and Ullrich, Sebastian},
  title     = {The {Lean} 4 Theorem Prover and Programming Language},
  booktitle = {Proceedings of the 28th International Conference on Automated Deduction},
  pages     = {625--635},
  year      = {2021},
  publisher = {Springer},
  doi       = {10.1007/978-3-030-79876-5_37},
}

@inproceedings{svcomp,
  title={Improvements in software verification and witness validation: SV-COMP 2025},
  author={Beyer, Dirk and Strej{\v{c}}ek, Jan},
  booktitle={International Conference on Tools and Algorithms for the Construction and Analysis of Systems},
  pages={151--186},
  year={2025},
  organization={Springer}
}

@software{rocq-lsp,
  author       = {Gallego Arias, Emilio Jesús and rocq-lsp contributors},
  title        = {Rocq-lsp: Visual Studio Code Extension and Language Server Protocol for Rocq},
  year         = {2025},
  version      = {v0.2.5},
  url          = {https://github.com/ejgallego/rocq-lsp},
  urldate      = {2025-11-30},
  note         = {GitHub repository}
}

@inproceedings{sel4,
  title={seL4: Formal verification of an OS kernel},
  author={Klein, Gerwin and Elphinstone, Kevin and Heiser, Gernot and Andronick, June and Cock, David and Derrin, Philip and Elkaduwe, Dhammika and Engelhardt, Kai and Kolanski, Rafal and Norrish, Michael and others},
  booktitle={Proceedings of the ACM SIGOPS 22nd symposium on Operating systems principles},
  pages={207--220},
  year={2009}
}

@inproceedings{rdbms,
  title={Toward a verified relational database management system},
  author={Malecha, Gregory and Morrisett, Greg and Shinnar, Avraham and Wisnesky, Ryan},
  booktitle={Proceedings of the 37th annual ACM SIGPLAN-SIGACT symposium on Principles of programming languages},
  pages={237--248},
  year={2010}
}

@inproceedings{compcert,
  title={CompCert-a formally verified optimizing compiler},
  author={Leroy, Xavier and Blazy, Sandrine and K{\"a}stner, Daniel and Schommer, Bernhard and Pister, Markus and Ferdinand, Christian},
  booktitle={ERTS 2016: Embedded Real Time Software and Systems, 8th European Congress},
  year={2016}
}

@article{pact,
  title={Proof artifact co-training for theorem proving with language models},
  author={Han, Jesse Michael and Rute, Jason and Wu, Yuhuai and Ayers, Edward W and Polu, Stanislas},
  journal={arXiv preprint arXiv:2102.06203},
  year={2021}
}

@inproceedings{deepseekprover1.5,
  title={Deepseek-Prover-v1. 5: Harnessing proof assistant feedback for reinforcement learning and monte-carlo tree search},
  author={Xin, Huajian and Ren, ZZ and Song, Junxiao and Shao, Zhihong and Zhao, Wanjia and Wang, Haocheng and Liu, Bo and Zhang, Liyue and Lu, Xuan and Du, Qiushi and others},
  booktitle={International Conference on Learning Representations},
  volume={2025},
  pages={72274--72303},
  year={2025}
}

@article{goedelprover,
  title={Goedel-prover: A frontier model for open-source automated theorem proving},
  author={Lin, Yong and Tang, Shange and Lyu, Bohan and Wu, Jiayun and Lin, Hongzhou and Yang, Kaiyu and Li, Jia and Xia, Mengzhou and Chen, Danqi and Arora, Sanjeev and others},
  journal={arXiv preprint arXiv:2502.07640},
  year={2025}
}

\end{document}